\documentclass[prl,amsmath,amssymb, notitlepage, twocolumn,
nofootinbib,
superscriptaddress,
longbibliography
]{revtex4-1}

\usepackage{amsmath}
\usepackage{tabularx,graphicx}
\usepackage{epstopdf}
\usepackage{graphicx}
\usepackage{latexsym}
\usepackage{amssymb}
\usepackage{amsmath}
\usepackage{color, colortbl}
\usepackage{psfrag}
\usepackage{bbm}
\usepackage{bm}
\usepackage{titlesec}
\usepackage{dsfont}
\usepackage{feynmp}
\usepackage{slashed}
\usepackage{multirow}
\newcommand{\mc}[1]{\mathcal{#1}}

\newcommand{\beq}{\begin{eqnarray}}
	\newcommand{\eeq}{\end{eqnarray}}

\newcommand{\la}{\langle}
\newcommand{\ra}{\rangle}

\newcommand{\bsp}{\begin{split}}
	\newcommand{\esp}{\end{split}}

\newcommand{\eff}{{\rm eff}}
\newcommand{\hc}{{\rm h.c.}}

\newcommand{\ie}{{i.e., }}
\newcommand{\eg}{{e.g., }}
\newcommand{\gcs}{{\rm{CS}}_g}

\usepackage{color}
\definecolor{darkblue}{rgb}{0.,0.,0.4}
\definecolor{darkred}{rgb}{0.5,0.,0.}
\definecolor{BlueViolet}{RGB}{138,43,226}
\definecolor{SkyBlue}{RGB}{30,144,255}
\definecolor{DarkGreen}{RGB}{0,100,0}
\usepackage[pdftex,colorlinks=true,linkcolor=darkblue,citecolor=blue,urlcolor=darkred]{hyperref}
\usepackage[normalem]{ulem}

\renewcommand{\vec}[1]{\bm{#1}}

\begin{document}
	\title{Deconfined Metal-Insulator Transitions in Quantum Hall Bilayers}
	
	\author{Liujun Zou}
	\affiliation{Perimeter Institute for Theoretical Physics, Waterloo, Ontario N2L 2Y5, Canada}
	
	\author{Debanjan Chowdhury}
	\affiliation{Department of Physics, Cornell University, Ithaca, New York 14853, USA}

\begin{abstract}
{\color{blue} } We propose that quantum Hall bilayers in the presence of a periodic potential at the scale of the magnetic length can host examples of a Deconfined Metal-Insulator Transition (DMIT), where a Fermi liquid (FL) metal with a generic electronic Fermi surface evolves into a gapped insulator (or, an insulator with Goldstone modes) through a continuous quantum phase transition. The transition can be accessed by tuning a single parameter, and its universal critical properties can be understood using a controlled framework. At the transition, the two layers are effectively decoupled, where each layer undergoes a continuous transition from a FL to a generalized composite Fermi liquid (gCFL). The thermodynamic and transport properties of the gCFL are similar to the usual CFL, while its spectral properties are qualitatively different. The FL-gCFL quantum critical point hosts a sharply defined Fermi surface without long-lived electronic quasiparticles. Immediately across the transition, the two layers of gCFL are unstable to forming an insulating phase. We discuss the topological properties of the insulator and various observable signatures associated with the DMIT.
\end{abstract}

\maketitle

{\it Introduction.} Understanding quantum criticality in metallic systems remains one of the outstanding challenges in the study of quantum matter, largely due to the abundance of gapless excitations near the electronic Fermi surface (FS). Describing an interaction-driven continuous metal-insulator transition is especially challenging, and a mechanism for an abrupt change of a generic electronic FS (at a fixed electron density) across a transition to an insulator in the absence of disorder is necessarily novel. Critical theories for continuous metal-insulator transitions, involving the disappearance of an entire electronic FS, have been formulated when the insulator is a quantum spin liquid with a FS of neutral excitations (\eg ``spinons'') coupled to emergent gauge fields \cite{TSFLstar,Senthil2008}. Pressure-tuned experiments on certain spin liquid candidates \cite{KKrev} have reported indirect evidence for such a transition \cite{KK15}. One of the most challenging and unsolved problems in the field is to describe a continuous metal-insulator transition where the insulator has {\it no} remnant FS of {\it any} excitations, which we dub a ``Deconfined Metal-Insulator Transition'' (DMIT), or a Deconfined Mott Transition in the terminology of Ref. \cite{Zou2020}. Such a transition necessarily falls outside the purview of the conventional Landau-Ginzburg-Wilson paradigm, even though the phases on either side of the transition can be conventional. Finding concrete theoretical examples of a DMIT in a solvable limit is therefore of paramount importance.

There are indirect signatures of continuous transitions between distinct metallic phases in numerous systems. For instance, in certain rare-earth element based (``heavy-fermion'') compounds there is evidence for entire Fermi surface sheets disappearing \cite{Stewart,Coleman00}, accompanied by a diverging effective mass near the putative critical point \cite{Shishido05}. Perhaps the most notable example of a similar transition arises in the cuprate superconductors \cite{Keimer15}, where a Fermi liquid (FL) metal evolves into an unconventional ``pseudogap'' metal  \cite{Badoux16}. It is important to note that none of these transitions can be described by the conventional framework of coupling an order-parameter field to an electronic FS. Recently, we have argued that the DMIT can be used as a building block to describe such experimentally relevant continuous transitions between distinct metallic phases, simply by including additional spectator electrons in the low-energy description that do not alter any fundamental aspects of the criticality itself  \cite{Zou2020}.

A general mechanism for DMIT based on ``emergent color superconductivity" \cite{Alford2007} was proposed in Ref. \cite{Zou2020}. In this work, inspired by the recent experimental advances in realizing quantum Hall (QH) physics in the presence of a periodic potential at the scale of the magnetic length (\eg in moir\' e heterostructures) \cite{PJH13,AKG13,PK13,AFY18}, we theoretically study a concrete setting for DMIT.

\begin{figure}[h]
\begin{center}
\includegraphics[scale=0.48]{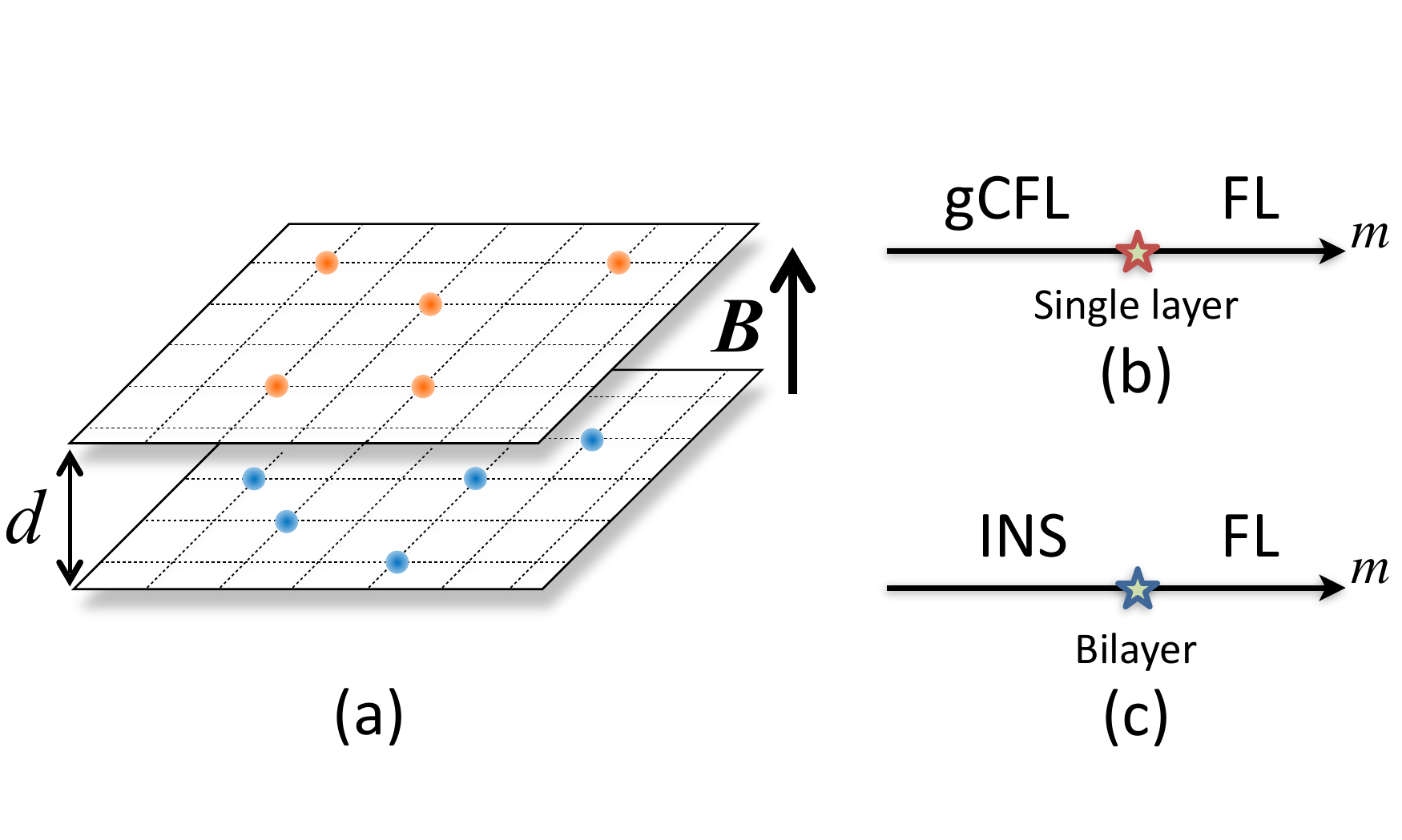}
\end{center}
\caption{(a) Schematic illustration of our proposed bilayer quantum Hall setup in an external $B$ field and a periodic potential. The layers are separated by a distance $d$. Phase diagram as a function of $m$ for (b) the single layer (or equivalently, two fully decoupled layers), and, (c) the bilayer. The FL-insulator (INS) transition is an example of a DMIT.}
\label{phasediag}
\end{figure}

{\it Setup and framework.} Consider a QH bilayer separated by a distance, $d$, with a periodic potential in each layer (Fig.~\ref{phasediag}); we will refer to this setup as a ``Chern bilayer''. To be concrete, we fix a unit flux quantum threading through each unit cell (UC) of the periodic potential. We will consider spinful electrons in each layer, such that spin-up electrons in the $S^z$-basis have a fixed density $\nu_\uparrow=2+\nu$ with $\nu=C/(C+1)$ per UC per layer ($C\in$ integer; $C\neq 0, -1$), while spin-down electrons have a fixed density $\nu_\downarrow=1$ per UC per layer. We will focus on $C>0$, but the phenomenology will be similar when $C<0$. We further assume that the electron number in each layer is conserved{\footnote{This assumption on particle number conservation in each layer can  be relaxed, as long as the total particle number is conserved.}}, the total $S^z$-spin is conserved, and the system is symmetric with respect to exchanging the two layers and with respect to an inversion within each layer. We are interested in the situation where the electrons interact via a repulsive two-body long-range density-density interaction $V(r)\sim 1/r^{1+\epsilon}$, with $r$ denoting the (three-dimensional) separation. The case with $\epsilon=0$ corresponds to Coulomb repulsion. 
We focus on $0<\epsilon<1$ for most of our discussion, and comment on the cases of $\epsilon=0$ and $\epsilon\geqslant 1$ at the end. We stress that the condition $0<\epsilon<1$ is introduced purely for technical convenience, and it should not be viewed as a fundamental limitation of our proposal. 

An equivalent description of this system is to view each layer as being made of spin-up electrons with fixed density $\nu'=1+\nu$ and spinless Cooper pairs with fixed density $\nu_B=1$. Below we will adopt this view since it is more convenient for our purpose. So each layer has ``valence'' (spin-up) electrons at a density of 1 per UC, and ``conduction'' (spin-up) electrons at a density of $\nu$. In fact, we will consider a scenario where the actions associated with the transitions mostly take place in the conduction electrons, while the valence electrons and Cooper pairs are gapped spectators that are coupled in a special way to enable the transitions to be direct. Our proposal will thus serve as a proof-of-concept setup for realizing DMIT in an electronic model.

To understand the DMIT in this setup, consider a single layer first. An earlier work \cite{Barkeshli2012} studied the possibility of a continuous transition between a FL and a composite Fermi liquid (CFL) of spinless electrons \cite{HLR}. However, it is our understanding that the resulting transition can only be accessed by tuning two parameters (instead of one) in order to avoid an intermediate phase. Interestingly, we will see that an analogous transition can be realized simply by tuning one parameter for the spinful electrons discussed above.
Building on this idea and the recent results of Ref. \cite{Lee2018} on transitions between distinct fractional Chern insulators, we can describe a continuous transition between a FL and a to-be-introduced ``generalized" CFL (gCFL) in our setup (Fig.~\ref{phasediag}b); the universal critical properties can be understood in a controlled manner when $C\gg 1$. For the bilayer (Fig.~\ref{phasediag}a) coupled via the long-range repulsion, we show that the corresponding gCFL phase is unstable to an insulator without FS of any excitations \cite{Bonesteel1996, Sodemann2016, Isobe2016, supp}. Moreover, when $C\gg 1$, we will show that the couplings between the two layers are renormalization-group (RG) irrelevant at the FL-gCFL transition. Therefore, our Chern bilayer hosts an example of a DMIT (Fig.~\ref{phasediag}c), where each layer undergoes a FL-gCFL transition (the latter being unstable to an insulator). Furthermore, the critical point hosts a non-Fermi liquid (nFL) with a sharply defined critical FS \cite{Senthil2008a}.
 
More specifically, denote the annihilation operator of spin-up electrons and spinless Cooper pairs on the $\ell$-th layer by $c_\ell$ and $B_\ell$, respectively. We express $c_\ell$ in terms of partons as $c_\ell=b_\ell f_\ell$, where $b_\ell$  and $f_\ell$ are bosonic and fermionic partons, respectively, that satisfy a constraint on their densities \cite{Wen2004Book},  $n_{b_\ell}=n_{f_\ell}=\nu'$. The low-energy theory for this parton construction of the spin-up electrons can be written in terms of the partons coupled to an emergent dynamical U$(1)$ gauge field, $a_\ell$. We let $b_\ell$ carry the global conserved U$(1)$ charge associated with the $\ell$-th layer and $f_\ell$ carry the spin (see Table \ref{tab: charge assignment main}), such that the schematic Lagrangian for the $\ell$-th layer takes the form:
\beq \label{eq: effective theory}
\mc{L}_\ell=\mc{L}_{[b_\ell, A_\ell-a_\ell]}+\mc{L}_{[f_\ell, a_\ell+A_s]}+\mathcal{L}_{[c_\ell, B_\ell]}+\cdots,
\eeq
where $A_\ell$ and $A_s$ are the probe gauge fields corresponding to the conserved charge on the $\ell$-th layer and total $S^z$-spin, respectively. The interlayer couplings will be discussed later. In the remainder of this paper we will take $\mc{L}_\ell$ to yield no net flux for $a_\ell$ and to describe a FS of $f_\ell$ at the mean-field level; we tune $\mc{L}_{[b_\ell, A_\ell-a_\ell]}$ to drive the transition. 

\begin{table}[h]
\setlength{\tabcolsep}{0.2cm}
\renewcommand{\arraystretch}{1.4}
    \centering
    \begin{tabular}{c|ccccc} 
    \hline \hline
    & $b_\ell$ & $f_\ell$ & $c_{\ell}$ & $B_\ell$ & $\psi_{\ell}$\\ \hline
    $A_{\ell}$ & 1 & 0 & 1 & 2 &0\\
    $A_s$ & 0 & 1 & 1 & 0 & 0\\
    $a_{\ell}$ & $-1$ & 1 & 0 & 0 & 0\\
    $\alpha_{\ell}$ & 0 & 0 & 0 & 0  &1\\
      \hline \hline
    \end{tabular}
    \caption{Charge assignment of the matter fields on the $\ell$-th layer under the different gauge fields.}
    \label{tab: charge assignment main}
\end{table}

One may wonder whether it requires fine-tuning to fix the flux of $a_\ell$ at the transition. We will see that in our setup the coupling between valence spin-up electrons and Cooper pairs, chosen as $\mc{L}_{[c_\ell, B_\ell]}=-\frac{2}{2\pi}A_\ell d(A_\ell-a_\ell)$, can fix the flux of $a_\ell$ without any fine-tuning. So below we will assume that $a_\ell$ has no net flux and focus on the conduction spin-up electrons. Later we will elaborate on the role of the gapped spectator valence electrons and Cooper pairs.

\begin{figure}[h]
\begin{center}
\includegraphics[scale=0.7]{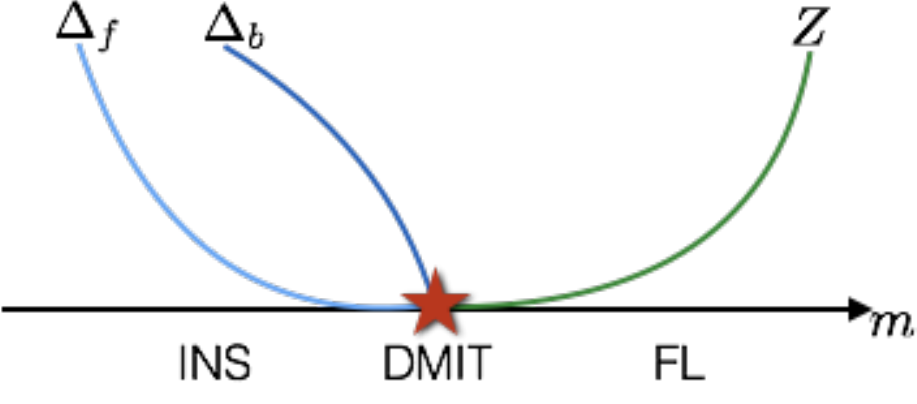}
\end{center}
\caption{A schematic diagram for the behavior of the quasiparticle residue, $Z$, the boson gap, $\Delta_b$, and the pairing gap for the fermions, $\Delta_f$, in the vicinity of the DMIT. The tuning parameter, $m$, appears in \eqref{eq: bosonic transition}.}
\label{fig: evolution}
\end{figure}

Before describing our results for the Chern bilayer in detail, we outline some key features associated with the phases and transitions of interest (see Fig.~\ref{fig: evolution}). By tuning a single parameter (denoted $m$ in Eq.~\eqref{eq: bosonic transition}), the bosonic partons can be driven from a superfluid (SF) to a QH state. In terms of electrons, the SF corresponds to a FL, and the SF condensate fraction determines the quasiparticle-residue, $Z$ (\ie the overlap between the electronic quasiparticle and the microscopic electron), which vanishes continuously upon approaching the DMIT. The boson gap, $\Delta_b$, opens up continuously on the QH side, where the latter corresponds to a gCFL in terms of electrons. One of the highlights of the present work is to identify a mechanism whereby, in the bilayer setting, the interlayer pairing of $f_\ell$ is {\it dangerously irrelevant}, leading to a continuous opening of the fermionic gap, $\Delta_f$. Remarkably, all three quantities vanish at the critical point, which hosts a sharply defined FS without any long-lived quasiparticles.

{\it Single-layer physics.} Consider a transition between a SF and a QH state for $b_\ell$, with the critical theory given by \cite{Barkeshli2012a, Lee2018, supp}
\beq \label{eq: bosonic transition}
\begin{split}
&\mc{L}_{[b_\ell, A_\ell-a_\ell]}
=\frac{(A_\ell-\alpha_\ell-a_\ell)d(A_\ell-\alpha_\ell-a_\ell)}{4\pi}+\gcs\\
&+\sum_{i=1}^{C+1}\overline\psi_{\ell i}(i\slashed  D_{\alpha_\ell}+m)\psi_{\ell i}
+\frac{(C-1)\alpha_\ell d\alpha_\ell}{8\pi}+\frac{(C-1)\gcs}{2},
\end{split}
\eeq
where $ada\equiv\epsilon_{\mu\nu\lambda}a_\mu\partial_\nu a_\lambda$, $\gcs$ is a gravitational Chern-Simons (CS) term \cite{Zou2018}, and ${\slashed D}_{\alpha_\ell}$ is the covariant derivative with respect to a {\it new} emergent gauge field, $\alpha_{\ell}$.
In this theory, $b_{\ell}^\dag$ is the monopole of $\alpha_\ell$ and the density fluctuations of $b_\ell$ correspond to the flux of $\alpha_\ell$; the transition is then driven by tuning the mass, $m$, of $(C+1)$ flavors of emergent Dirac fermions, $\psi_{\ell i}$. \footnote{In terms of the flux-attachment picture, the first term in Eq.~(\ref{eq: bosonic transition}) attaches one flux quantum to each boson $b_\ell$ and converts it into a fermion, $\psi_\ell$, which thereby sees a net flux of $1-\nu=\frac{1}{C+1}$ per UC. The phase transition corresponds to one where the Chern number of $\psi_\ell$ changes from $-1$ to $C$, which is captured \cite{Lee2018} by the last three terms in Eq.~(\ref{eq: bosonic transition}). An alternative derivation for Eq.~(\ref{eq: bosonic transition}) appears in \cite{supp}. }

What are the different phases that the theory given by \eqref{eq: effective theory} and \eqref{eq: bosonic transition} can describe by varying $m$? When $m>0$, for any $C$, integrating $\psi_{\ell i}$ out removes the CS term of $\alpha_\ell$ and the resulting state is a SF of $b_\ell$, as is evident from the
Dasgupta-Halperin duality \cite{Dasgupta1981}. In terms of the original $c_\ell$, this is just a FL phase when $f_\ell$ forms a FS. {\footnote{When discussing the various phases, we have assumed that the valence part of $f_\ell$ has no Chern number. It is straightforward to generalize the discussion to the case where it has a Chern number. Notice that the universal critical physics at the transitions does not depend on this Chern number.}} On the other hand, when $m<0$, integrating $\psi_{\ell i}$ out leads to
\beq \label{eq: bosonic QH}
\begin{split}
\mc{L}_{[b_\ell, A_\ell-a_\ell]}
&=
\frac{C+1}{4\pi}\alpha_\ell ~d\alpha_\ell + \frac{1}{2\pi}(a_\ell-A_\ell)~d\alpha_\ell\\
&+\frac{1}{4\pi}(a_\ell-A_\ell)~d(a_\ell-A_\ell)+(C+1)\gcs,
\end{split}
\eeq
which is a QH state of $b_\ell$. When $f_\ell$ forms a FS, the resulting state of the original $c_\ell$ is a gCFL \cite{supp}. Indeed, upon further integrating out the gapped $\alpha_\ell$, the effective theory reduces to
\beq \label{eq: gCFL effective theory}
\begin{split}
\mc{L}_{\ell}
=&\mc{L}_{[f_\ell, a_\ell+A_s]}
+\frac{\nu}{4\pi}(a_\ell-A_\ell)d(a_\ell-A_\ell)+\mc{L}_{[c_\ell, B_\ell]}.
\end{split}
\eeq
When $C=1$ and ignoring $\mc{L}_{[c_\ell, B_\ell]}$, this is precisely the effective theory of the familiar CFL at $\nu=1/2$, which hosts a nFL of $f_\ell$ \cite{HLR}. When $C\in$ integers ($\neq0, \pm 1$), the qualitative picture for the universal physics associated with $(f_\ell, a_\ell)$ largely remains the same, in that the thermodynamic and transport properties of the gCFL are similar to the usual CFL, while the spectral properties are qualitatively different \cite{supp}.

Next we turn to the critical point in Eq.~\eqref{eq: bosonic transition}, and ignore its coupling to the $(f_\ell, a_\ell)$-sector (and the spectator-sector). An important observation is that, for any $C$, the magnetic translation symmetry forbids various potentially relevant perturbations at this transition, such that the transition can be accessed by tuning a single parameter, $m$ \cite{Lee2018}. When $C\gg 1$, this theory can be studied in a controlled manner, and it flows in the IR to a conformal field theory (CFT) \cite{Chen1993}. As shown in Fig. \ref{fig: evolution} (but for the single-layer problem), $Z\sim|\la b_\ell\ra|^2$ and $\Delta_b$ vanish continuously as we approach the transition from the superfluid and QH side, respectively; $\Delta_f=0$ in the present single-layer setup.

Note that the long-range interaction $V(r)$ is not included above, which takes the form (in the Coulomb gauge with $\vec\nabla\cdot\vec\alpha_\ell=0$),
\beq \label{eq: long-range potential}
\int_{\omega,\vec k}k^{1+\epsilon}\alpha_{\ell t}(-\vec k, -\omega)\alpha_{\ell t}(\vec k, \omega),
\eeq
with $\alpha_{\ell t}$ the transverse spatial component of $\alpha_\ell$ \cite{HLR}. When $\epsilon>0$, this interaction is irrelevant compared to the CS term of $\alpha_\ell$, so it can be ignored at the critical point. In passing, we note that a Maxwell term for $\alpha_\ell$ is invariably generated due to local interactions, which is less (equally  or more) relevant than \eqref{eq: long-range potential} if $\epsilon<1$ ($\epsilon\geqslant 1$). So this long-range interaction is important if $\epsilon<1$, while if $\epsilon\geqslant 1$, we can ignore it and consider only local interactions as far as universal critical physics is concerned.

Now we need to combine the $b_\ell$-sector and the $(f_\ell, a_\ell)$-sector, which are coupled via operators of the form $\mc{O}_{b_\ell}\mc{O}_{(f_\ell, a_\ell)}$, where $\mc{O}_{b_\ell}$ and $\mc{O}_{(f_\ell, a_\ell)}$ are gauge invariant operators from the two sectors, respectively. Naively, as long as the scaling dimensions of these operators are large enough, such couplings are RG irrelevant and the two sectors are effectively decoupled. Indeed, as pointed out in Ref. \cite{Senthil2008}, when the scaling dimensions of all gauge invariant operators in the $b_\ell$-sector are larger than $3/2$, the criticality of $b_\ell$ is unaffected by the presence of the $(f_\ell, a_\ell)$-sector. When $C\gg 1$, this condition is satisfied for the CFT corresponding to \eqref{eq: bosonic transition} \cite{Chen1993}. However, the presence of the critical $b_\ell$-sector has a significant influence on the dynamics of the $(f_\ell, a_\ell)$-sector. In particular, integrating out $b_\ell$ at the critical point generates the following effective action for $a_\ell$:
\beq \label{eq: effective a due to critical b}
\delta S_{a_\ell}=\sigma_b\int_{\omega, \vec k}\sqrt{\omega^2+k^2}~|a_{\ell t}(\vec k, \omega)|^2,
\eeq
with $\sigma_b$ a universal constant determined by the CFT corresponding to \eqref{eq: bosonic transition}, $a_{\ell t}$ the transverse spatial component of $a_\ell$, and the boson velocity is set to unity. As a result of Landau-damping due to coupling to the FS of $f_\ell$, the effective action for $a_{\ell t}$ has an additional contribution, $\int_{\omega, \vec k}\frac{|\omega|}{k}|a_{\ell t}(\vec k, \omega)|^2$, leading to a dynamical exponent $z_a=2$. As a result, at the critical point, the $f_\ell$ are endowed with a marginal FL-like self-energy and the electrons have a well-defined critical FS without long-lived quasiparticles.

Finally, we examine the important and subtle issue of the flux of $a_\ell$ and the role of the spectator valence electrons and Cooper pairs. If there is no nontrivial coupling $\mc{L}_{[c_\ell, B_\ell]}$ in \eqref{eq: effective theory}, nothing fixes this flux at the transition. So without fine-tuning, a net flux of $a_\ell$ can be generated and the critical point is expected to be unstable to forming certain QH state. However, unlike earlier approaches \cite{Barkeshli2012}, this aspect can be circumvented by incorporating a nontrivial $\mc{L}_{[c_\ell, B_\ell]}$. Specifically, denote the valence part of $b_\ell$ by $b_{\ell v}$, and attach a vortex of $B_\ell$ to $b_{\ell, v}$, and a vortex of $b_{\ell, v}$ to $B_\ell$. This coupling induces an interaction $\mc{L}_{[c_\ell, B_\ell]}=-\frac{2}{2\pi}A_\ell d(A_\ell-a_\ell)$ at low energies \cite{Senthil2012}, which means the flux of $a_\ell$ now carries charge under $A_\ell$. Because the charge under $A_\ell$ is fixed, now the flux of $a_\ell$ is also fixed (without fine-tuning) \cite{supp}.

{\it DMIT in bilayer.} When the two identical layers with separately conserved densities are coupled through the long-range repulsion, we obtain two effectively decoupled FL for $m>0$. What is the fate of the system when $m<0$?

A useful starting point is to consider the limit where the two layers are completely decoupled so that each of them can be captured within the above discussion, and then study the effect of interlayer couplings as we tune through the individual single-layer FL-gCFL transition. This scenario can be physically realized when $d$ is large.

We begin by introducing $a_\pm\equiv (a_1\pm a_2)/2$, where the layer-exchange symmetry forbids a direct coupling between $a_+$ and $a_-$. As argued in Refs. \cite{Bonesteel1996, Zou2016, Sodemann2016, Isobe2016, Zou2020}, $a_+$ tends to {\it suppress} pairing of the FS while $a_-$ {\it favors} interlayer pairing, such that their competition determines the stability of the FS. When $b_\ell$ is gapped, \eqref{eq: gCFL effective theory} indicates that the density fluctuation of $f_\ell$ is related to the flux of $a_\ell$, so the long-range repulsion becomes an interaction between flux of $a_\pm$. Notice that $a_+$ couples to the total density of $f_1$ and $f_2$, while $a_-$ couples to the density difference between $f_1$ and $f_2$. So the flux of $a_+$ experiences the long-range potential, while the interaction between the flux of $a_-$ is effectively short-ranged. 
As a result, the coupling between $a_-$ and the fermions is more relevant and the FS is unstable to interlayer pairing \cite{supp}.  This implies that two layers of gCFL are unstable to forming an insulating phase and we discuss its topological character shortly. Since the boson gap, $\Delta_b$, serves as the effective UV cutoff of the low-energy physics associated with the interlayer pairing, the resulting pairing gap, $\Delta_f$, should be smaller than $\Delta_b$ (see Fig. \ref{fig: evolution}).

At the critical point, we should instead use the effective action in \eqref{eq: effective a due to critical b} to describe the gauge fields and determine the stability of the FS. Rewriting
\beq
\sum_\ell\delta S_{a_\ell}=\int_{\omega, \vec k}\sqrt{\omega^2+k^2}\left[\frac{|a_{+t}(\vec k, \omega)|^2}{2g_+}+\frac{|a_{-t}(\vec k, \omega)|^2}{2g_-}\right],\nonumber\\
\eeq
with $a_{\pm t}$ the transverse spatial component of $a_\pm$, we can identify the gauge couplings $g_+=g_-=1/(4\sigma_b)$. When $z_a=2$, the FS is perturbatively stable if $g_+\geqslant g_-$ \cite{Zou2020}.{\footnote{Kohn-Luttinger type effects \cite{Kohn1965, Shankar1994} are ignored.}} Therefore, interlayer pairing is (dangerously) irrelevant at this transition.

Can other interlayer couplings alter the critical properties? First, just as in the case of the monolayer, the interlayer long-range interaction is irrelevant at the critical point if $\epsilon>0$. In addition, there are local interactions that all turn out to be irrelevant \cite{Zou2016, supp}. For example, there can be a coupling of the form $\mc{O}_{b_1}\mc{O}'_{b_2}$, where $\mc{O}$ and $\mc{O}'$ are gauge invariant operators in the $b_1$- and $b_2$-sector, respectively. Clearly this coupling is irrelevant if the scaling dimensions of $\mc{O}$ and $\mc{O}'$ are larger than $3/2$. In fact, a sufficient condition for the layer decoupling is that all gauge invariant operators of the $b_\ell$-sector have a scaling dimension larger than $3/2$ \cite{supp}, a condition satisfied when $C\gg 1$ \cite{Chen1993}.{\footnote{Although we mainly focus on the case where the particle density in each layer is separately conserved, our conclusions regarding the interlayer pairing instability of $f$ and the DMIT are robust even in the presence of a weak interlayer electron tunneling, which leads to a single conserved U(1) density, and is irrelevant at the DMIT \cite{supp}.}}

We have thus reached the remarkable conclusion that the above bilayer setup can exhibit a DMIT, where the critical point hosts two effectively decoupled, sharp critical FS.

{\it Topological properties of the insulating phase.} The effective theory for the insulating phase in the bilayer setting is given by $\sum_\ell\mc{L}_{\ell}$ in (\ref{eq: effective theory}), with $\mc{L}_{[b_\ell, A_\ell-a_\ell]}$ given by (\ref{eq: bosonic QH}), $\mc{L}_{[f, a+A_s]}$ the effective Lagrangian for the interlayer-paired state of the fermions, and $\mc{L}_{[c_\ell, B_\ell]}=-\frac{2}{2\pi}A_\ell d(A_\ell-a_\ell)$. Our goal is to characterize this phase via the $K$-matrix formalism \cite{Wen2004Book}.

Since $\alpha_\ell$ in (\ref{eq: bosonic QH}) is coupled to a fermion, to apply the standard $K$-matrix formalism, we introduce $C+1$ gauge fields, $\beta_{\ell}^i$ ($i=1, 2, \cdots, C+1$) for the $\ell$-th layer, which are coupled to bosons.  In terms of these new gauge fields, $\mc{L}_{[b_\ell, A_\ell-a_\ell]}$ is equivalent to:
\beq \label{eq: Bosonic QH equivalent}
\begin{split}
\mc{L}_{[b_\ell, A_\ell-a_\ell]}
&=-\frac{1}{4\pi}\sum_{i=1}^{C+1}\beta_{\ell}^i~d\beta_{\ell}^i-\frac{\alpha_\ell}{2\pi}\sum_{i=1}^{C+1}d\beta_{\ell}^i\\
+&\frac{1}{2\pi}(a_\ell-A_\ell)~d\alpha_\ell+\frac{1}{4\pi}(a_\ell-A_\ell)~d(a_\ell-A_\ell).
\end{split}
\eeq
Note that integrating out $\beta_{\ell}^i$'s above reproduces (\ref{eq: bosonic QH}). We now integrate out $\alpha_\ell$ and obtain a constraint \beq \label{eq: elimination}
\sum_{i=1}^{C+1}\beta_{\ell}^i=a_\ell-A_\ell.
\eeq

Next we turn to $\mc{L}_{[f, a+A_s]}$. The channel in which interlayer pairing occurs depends on non-universal details \cite{Bonesteel1996, Sodemann2016, Isobe2016, supp}. Suppose it occurs in the channel with index $n$ in Kitaev's 8-fold way \cite{Kitaev2006}. {\footnote{Since our interlayer pairing is between two species of fermions, only the Abelian 8-fold way of classification is relevant here among the full 16-fold way classification of Kitaev, so we use a convention that our $n$ is half of Kitaev's index in Ref. \cite{Kitaev2006}.} The topological nature of the paired state depends on $n$, which can be described with the introduction of low-energy emergent gauge fields in addition to $a$ \cite{Kitaev2006}. Naturally, it is important to understand how these new gauge fields couple to $a_\pm$ and $A_s$.

A convenient way to proceed is to separate the charges under $a_++A_s$ and $a_-$ by performing a parton decomposition of $f_\ell\equiv\phi d_\ell$, where $\phi$ is a boson such that $\phi^2$ is the interlayer pairing amplitude, $d_\ell$ is a fermion, such that $\phi$ and $d_\ell$ are coupled to an emergent $Z_2$ gauge field.
Here $\phi$ carries charge $1$ under $a_++A_s$ but no charge under $a_-$, and $d_1$ and $d_2$ carry charge 1 and $-1$ under $a_-$, respectively, but are neutral under $a_++A_s$. In the interlayer-paired state of $f_\ell$, $\phi$ is condensed and $d_\ell$ develops interlayer pairing in the same channel as $f_\ell$. The condensate of $\phi$ can be captured by $\frac{1}{\pi}(a_++A_s)d\beta$, where $\beta$ is a new emergent gauge field and its elementary charge binds the $\pi$-flux of the $Z_2$ gauge field. This binding is why the coefficient of $(a_++A_s)d\beta$ is $1/\pi$, not $1/(2\pi)$  \cite{Senthil2000}. 

The $d_\ell$-sector depends on $n$, and it is most convenient to start with the case $n=\pm 1$. It is known that when the state with $n=\pm 1$ is coupled to a dynamical $Z_2$ gauge field, the resulting state is U$(1)_{\pm 4}$, captured by $\mp\frac{4}{4\pi}\gamma d\gamma$, where odd charge of the new emergent gauge field $\gamma$ is identified with the $\pi$-flux of the $Z_2$ gauge field \cite{Kitaev2006}. This identification further constrains that $q_\beta=q_\gamma\ ({\rm mod\ } 2)$, where $q_{(\cdot)}$ represents the possible charge that an excitation carries under the corresponding gauge field. Since $d_\ell$ is coupled to $a_-$, we need to determine how $\gamma$ is coupled to $a_-$. The layer exchange symmetry requires that they are coupled via $\frac{1}{\pi}a_-d\gamma$ \cite{Sodemann2016}. Therefore, when $n=\pm 1$,
\beq \label{eq: |nu|=1}
\mc{L}_{[f, a+A_s]}=\frac{1}{\pi}(a_++A_s)d\beta\mp\frac{4}{4\pi}\gamma d\gamma+\frac{1}{\pi}a_-d\gamma
\eeq
with a constraint $q_\beta=q_\gamma\ ({\rm mod\ } 2)$. To apply the standard $K$-matrix formalism, we can solve this constraint by introducing $\tilde\beta\equiv\beta+\gamma$ and $\tilde\gamma\equiv\beta-\gamma$. The charges of $\tilde\beta$ and $\tilde\gamma$ can independently take any integer. Denote the electric and spin Hall conductivity by $\sigma_{xy}^c$ and $\sigma_{xy}^s$, respectively. Plugging \eqref{eq: bosonic QH}, \eqref{eq: |nu|=1} and $\mc{L}_{[c_\ell, B_\ell]}=-\frac{2}{2\pi}A_\ell d(A_\ell-a_\ell)$ into $\sum_\ell\mc{L}_\ell$ yields $\sigma_{xy}^{c}=2\nu-8$ and $\sigma_{xy}^s=2\nu$.

To fully unearth the topological property, we write down the final $K$-matrix theory for $n=\pm 1$ in terms of $\tilde a\equiv(\beta_{1}^1, \beta_{1}^2, \cdots, \beta_{1}^{C+1}, \beta_{2}^1, \beta_{2}^2, \cdots, \beta_{2}^{C+1}, \tilde\beta, \tilde\gamma)^T$:
\beq
\mc{L}=\frac{K_{IJ}}{4\pi}\tilde a_Id\tilde a_J+\frac{t_{1I}}{2\pi}A_1d\tilde a_I+\frac{t_{2I}}{2\pi}A_2d\tilde a_I+\frac{t_{sI}}{2\pi}A_sd\tilde a_I
\eeq
with
\beq \label{eq: K-matrix}
\begin{split}
&K=\left(
\begin{array}{ccc}
K_1 & 0 & K_2\\
0 & K_1 & K_3\\
K_2^T & K_3^T & K_4
\end{array}
\right)\\
&t_1=(2, 2, \cdots, 2,0, 0, \cdots, 0, 1, 0)^T\\
&t_2=(0, 0, \cdots,0, 2, 2, \cdots, 2, 0, 1)^T\\
&t_s=(0,0,\cdots, 0, 1, 1)^T
\end{split}
\eeq
Here $K_1$ is a $(C+1)\times (C+1)$ matrix whose diagonal elements vanish while all other entries are 1, $K_2$ ($K_3$) is a $(C+1)\times 2$ matrix whose entries in the first (second) column are all 1 while all other entries vanish, and
\beq
K_4=
\left(
\begin{array}{cc}
\mp 1 & \pm 1\\
\pm 1 & \mp 1
\end{array}
\right).
\eeq
The first $C+1$ entries of $t_1$ ($t_2$) is 2 (0). In general, the above theory describes a topological order, but when $C=1$ and $n=-1$, the resulting state has a Goldstone mode due to the spontaneous breaking of a U(1) symmetry generated by $5(Q_1-Q_2)-S^z$, with $Q_{1,2}$ the electric charge of the two layers, respectively \cite{supp}.

To obtain the $K$-matrix for states with other $n\neq\pm1$ systematically, the simplest approach may be to apply the trick in Ref. \cite{Zou2018} to build up the theory from that with $n=\pm 1$ \cite{supp}.

{\it Observable signatures.} The DMIT can be viewed as two dynamically decoupled FL-gCFL transitions (where the gCFL-bilayer is unstable to an insulator). Therefore, some of the universal physical properties at both transitions are similar \cite{Barkeshli2012, Lee2018}, which include: (i) a critical FS, (ii) a singular specific heat ($\sim T\ln T$) at the transition, (iii) a jump of the electric resistivity ($\sim h/e^2$), (vi) an emergent SU($C+1$) symmetry where a set of power-law-decaying charge-density-wave order parameters transform in its adjoint representation. One main difference between the monolayer and bilayer systems appears in the insulating side of $b_\ell$, where the bilayer shows interlayer pairing of $f_\ell$ below certain temperature scale \cite{Bonesteel1996, Sodemann2016, Isobe2016}.

{\it Outlook.} In this work, we have primarily focused on the case of a repulsive two-body interaction, $V(r)\sim 1/r^{1+\epsilon}$, with $0<\epsilon<1$, at the DMIT. For the case of usual Coulomb repulsion ($\epsilon=0$), the interaction is marginal at the tree-level with respect to the bosonic superfluid-QH transition. We leave a detailed study on the effect of Coulomb interaction on such transitions for the future \cite{Ye1997}. The case with $\epsilon\geqslant 1$ may be realized with cold atoms \cite{Cooper2013, Yao2013}, and the physics in this case reduces to that in Ref. \cite{Zou2020}. The crossovers at finite temperature out of the regimes considered here are expected to be rich and we leave a detailed study of this phenomenology for future work, along with a study of the effects of different types of disorder on these transitions. The transport properties in the quantum critical regime associated with the DMIT will likely shed interesting light on the non-Fermi liquid. It would be interesting to find concrete models for these DMIT and study them numerically. A smoking-gun signature for the DMIT in the density-density response would be a sharp ``$2K_F$" response arising from the critical Fermi surface, in addition to the set of isolated peaks from the charge-density wave due to the emergent SU$(C+1)$ symmetry at the critical point.

{\it Acknowledgement.} We thank Zhen Bi, Inti Sodemann and especially Chong Wang for useful discussions. LZ is supported by the John Bardeen Postdoctoral Fellowship at Perimeter Institute. Research at Perimeter Institute is supported in part by the Government of Canada through the Department of Innovation, Science and Economic Development Canada and by the Province of Ontario through the Ministry of Colleges and Universities. DC is supported by faculty startup funds at Cornell University.

\bibliography{Chernbilayer.bib}

\clearpage
\onecolumngrid
\appendix
\begin{center}
	{\bf Supplementary Material for ``Deconfined Metal-Insulator Transitions in Quantum Hall Bilayers"}
\end{center}
\vspace{0.5cm}

This Supplementary Material contains additional details on: {\rm I.} the critical theory of the transition between the quantum Hall (QH) state and superfluid of the bosons, {\rm II.} some of the low-energy properties of the generalized composite Fermi liquid (gCFL) phase, {\rm III.} additional details on the spectator valence electrons and Cooper pairs, {\rm IV.} a discussion of the interlayer pairing instability of two layers of gCFL, {\rm V.} an analysis of the effects of various interlayer couplings at the metal-insulator transition, and, {\rm VI.} details of deriving the topological properties of the insulating phases.

\section{Bosonic quantum Hall - superfluid transition}

In the main text we have provided a flux-attachment based picture for the single-layer QH - superfluid transition of the ``conduction'' bosonic partons derived from the spin-up electrons.
In this section we provide an alternative interpretation of the same critical theory, and we only focus on the conduction bosons with density $\nu=C/(C+1)$ per UC per layer.

In order to describe the QH state of $b_{\ell}$ and the associated transition into a superfluid of $b_\ell$, we introduce two more fermionic partons $\chi_{\ell}$ and $\psi_{\ell}$, such that $b_\ell=\chi_{\ell}\psi_{\ell}$, supplemented with a constraint $n_{\chi_{\ell}}=n_{\psi_{\ell}}=n_{b_\ell}=\nu$. This parton construction introduces an SU(2) gauge redundancy, but we will explicitly break it down to U$(1)$ \cite{Wen2004Book}. This means that at low energies the $b_\ell$-sector can be described by $\chi_{\ell}$ and $\psi_{\ell}$ coupled to another emergent dynamical U$(1)$ gauge field, $\alpha_\ell$. The charge assignment for the different partons under all of the gauge fields is summarized in Table \ref{tab: charge assignment}. 

\begin{table}[h]
	\setlength{\tabcolsep}{0.2cm}
	\renewcommand{\arraystretch}{1.4}
	\centering
	\begin{tabular}{c|ccccc} 
		\hline \hline
		& $b_\ell$ & $f_\ell$ & $c_\ell$ & $\chi_{\ell}$ & $\psi_{\ell}$\\ \hline
		$A_{\ell}$ & 1 & 0 & 1 & 1 &0\\
		$A_s$ & 0 & 1 & 1 & 0 & 0\\
		$a_{\ell}$ & $-1$ & 1 & 0 & $-1$ & 0\\
		$\alpha_{\ell}$ & 0 & 0 & 0 & $-1$ &1\\
		\hline \hline
	\end{tabular}
	\caption{Charge assignment of the partons under the different gauge fields.}
	\label{tab: charge assignment}
\end{table}

The theory for the $b_\ell$-sector can then be written as
\beq
\mc{L}_{[b_\ell, A_\ell-a_\ell]}=\mc{L}_{[\chi_{\ell}, -\alpha_\ell-a_\ell+A_\ell]}+\mc{L}_{[\psi_{\ell}, \alpha_\ell]}.
\eeq
Notice in discussing this bosonic sector, $A_\ell-a_\ell$ is treated as a static gauge field.

Consider now a mean field state where $a_\ell$ has no average flux and $\alpha_\ell$ has an average of $1/(C+1)~(\equiv 1-\nu)$ flux quanta per UC. Using the charge assignment in Table \ref{tab: charge assignment}, this means that, modulo the unit flux quantum per UC  experienced by $b_\ell$ from the external gauge field, $f_\ell$ and $b_\ell$ experience no flux, $\chi_{\ell}$ experiences a total flux of $C/(C+1)~(\equiv \nu)$ flux quanta per UC, and $\psi_{\ell}$ experiences $1/(C+1)$ of flux quanta per UC. Then with particle density $n_{c_\ell}=n_{b_\ell}=n_{\chi_{\ell}}=n_{\psi_{\ell}}=C/(C+1)$, we can put $\chi_{\ell}$ into a mean field state with Chern number $1$, and tune the parameters of the system so that $\psi_{\ell}$ undergoes a transition from a state with Chern number $-1$ to a state with Chern number $C$ \cite{Lee2018, Barkeshli2012a}. The critical theory of this transition can be described by \beq \label{eq: bosonic transition supp}
\begin{split}
	\mc{L}_{[b_\ell, A_\ell-a_\ell]}
	=&\frac{1}{4\pi}(A_\ell-\alpha_\ell-a_\ell)d(A_\ell-\alpha_\ell-a_\ell)+\gcs\\
	+&\frac{C-1}{8\pi}\alpha_\ell d\alpha_\ell+\frac{(C-1)\gcs}{2}+\sum_{i=1}^{C+1}\overline\psi_{\ell, i}(i\slashed  D_{\alpha_\ell}+m)\psi_{\ell, i},
\end{split}
\eeq
where the $C+1$ flavors of Dirac fermions $\psi_{\ell, i}$ are the low-energy modes of $\psi_{\ell}$. This is precisely the critical theory in the main text. Notice $b_\ell^\dag$ is identified as the monopole of $\alpha_\ell$ in this theory, since the $2\pi$-flux of $\alpha_\ell$ carries charge $-1$ under $A_\ell-a_\ell$. Also, this theory has an emergent SU$(C+1)$ flavor symmetry, under which the Dirac fermions transform in the fundamental representation. Because of the Chern-Simons term of $\alpha_\ell$, the monopole of $\alpha_\ell$ that is neutral under $\alpha_\ell$ has no zero mode filled, which implies that $b_\ell$ is a singlet under the emergent SU$(C+1)$ symmetry \cite{Borokhov2002}.

In the above effective Lagrangian, the first two terms represent $\chi_{\ell}$ that is in a state with Chern number 1, and the last three terms represent the state of $\psi_{\ell}$. If $m>0$, integrating out $\psi_{\ell, i}$ converts the last three terms into
\beq
\mc{L}_{\ell 2}=-\frac{1}{4\pi}\alpha_\ell~ d\alpha_\ell-\gcs,
\eeq
which indeed describes a state of $\psi_{\ell}$ that has Chern number $-1$. Combining it with the first two terms, we obtain
\beq
\mc{L}_{[b_\ell, A_\ell-a_\ell]}=\frac{1}{4\pi}(A_\ell-a_\ell)~d(A_\ell-a_\ell)-\frac{1}{2\pi}(A_\ell-a_\ell) ~d\alpha_\ell.
\eeq
Notice the absence of a CS term for $\alpha$, so this is a superfluid state of $b_\ell$ \cite{Dasgupta1981}. As $b_\ell$ transforms trivially under all global symmetries, this superfluid does not further spontaneously break other global symmetries.

If $m<0$, integrating out $\psi_\ell$ converts the last three terms of (\ref{eq: bosonic transition supp}) into
\beq
\mc{L}_{\ell 2}=\frac{C}{4\pi}\alpha_\ell~ d\alpha_\ell + C\cdot \gcs,
\eeq
which describes a state of $\psi_{\ell}$ that has Chern number $C$. Combining it with the first two terms in (\ref{eq: bosonic transition supp}), we obtain the effective theory of the QH state of $b_\ell$ with a Hall conductivity $\sigma_{xy}^{(b)}=\nu=C/(C+1)$:
\beq \label{eq: boson QH effective supp}
\mc{L}_{[b_\ell, A_\ell-a_\ell]}=\frac{C+1}{4\pi}\alpha_\ell ~d\alpha_\ell + \frac{1}{2\pi}(a_\ell-A_\ell)~d\alpha_\ell
+\frac{1}{4\pi}(a_\ell-A_\ell)~d(a_\ell-A_\ell)+(C+1)\gcs.
\eeq

\section{ Low-energy properties of the gCFL with $C\neq 1$}

In this section, we discuss details of the gCFL state with $C\neq 1$. We will argue that this state can be viewed as a QH-version of an orthogonal metal (OM) \cite{Nandkishore2012}, as qualitatively it has similar thermodynamic and transport properties as the usual spinless CFL \cite{HLR}, but the spectral property is markedly different. In particular, in the usual CFL, the single electron has a soft gap \cite{BH93,Kim1994}. However, in the gCFL the single electron has a hard gap, while only multi-electron bound states have soft gaps. So the analogy between the usual CFL and the gCFL is analogous to the analogy between the usual Fermi liquid and an OM. 

First, we remark that the construction of the gCFL involves the procedure of ``flux removal" in the conduction sector, which removes the flux from the conduction electron $c_\ell$ to the boson $b_\ell$, so that $f_\ell$ can form a Fermi surface and $b_\ell$ can form a QH state. Flux removal is the real essence of construction of CFL \cite{HLR}, while the usually-stated ``flux attachment" is just a particular way to implement flux removal.

To understand the thermodynamic and transport properties of the gCFL, one can apply the Ioffe-Larkin-type analysis \cite{Ioffe1989}, with the QH physics and the spectator sector taken into account. Specifically, recall that the effective theory reads
\beq
\mc{L}_\ell=\mc{L}_{[b_\ell, A_\ell-a_\ell]}+\mc{L}_{[f_\ell, a_\ell+A_s]}-\frac{1}{\pi}A_\ell d(A_\ell-a_\ell)
\eeq
For simplicity, we will set $A_s=0$ here. Integrating out $b_\ell$ and $f_\ell$ results in the following effecitve Lagrangian
\beq
\begin{split}
	\mc{L}_\eff=&\frac{1}{2}\left[A_\ell(k)-a_\ell(k)\right]^T\Pi_b(k)\left[A_\ell(-k)-a_\ell(-k)\right]
	+\frac{1}{2}a_\ell(k)^T\Pi_f(k)a_\ell(-k)\\
	+&\frac{1}{2}\left[A_\ell(k)^T\Pi_s(k)a_\ell(-k)+a_\ell(k)^T\Pi_s(k)A_\ell(-k)\right]-A_\ell(k)^T\Pi_s(k)A_\ell(-k)
\end{split}
\eeq
where $k=(\omega, \vec q)$ compactly denotes the frequency and wavevector, $\Pi_{b}$ and $\Pi_{f}$ represent the polarization tensors of $b_\ell$ and $f_\ell$, respectively, and $\Pi_s$ is the contribution from the spectator sector. We will take the Coulomb gauge, \ie $\vec\nabla\cdot\vec A=\vec\nabla\cdot\vec a=0$, so the gauge fields are 2-component vectors, \eg $a_\ell=(a_{\ell 0}, a_{\ell t})$, with $a_{\ell 0}$ and $a_{\ell t}$ the temporal and transverse spatial component of $a_\ell$, respectively, such that $a_{\ell 1}(k)=\frac{q_2}{q}a_{\ell t}(k)$ and $a_{\ell 2}(k)=-\frac{q_1}{q}a_{\ell t}(k)$. In this gauge,
\beq \label{eq: polarization spectator}
\Pi_s(k)=
\left(
\begin{array}{cc}
	0 & -\frac{q}{\pi} \\
	\frac{q}{\pi} & 0
\end{array}
\right)
\eeq
Deep in the QH state of $b_\ell$, $\Pi_b$ takes the form
\beq \label{eq: polarization boson}
\Pi_b(k)=
\left(
\begin{array}{cc}
	0 & -\frac{\sigma_{xy}^{(b)} q}{2\pi} \\
	\frac{\sigma_{xy}^{(b)} q}{2\pi} & 0
\end{array}
\right)
\eeq
where $\sigma_{xy}^{(b)}=\nu$ is the Hall conductivity of the conduction bosons. And since $f_\ell$ has a Fermi surface, $\Pi_f$ takes the form
\beq \label{eq: polarization fermion}
\Pi_f(k)=
\left(
\begin{array}{cc}
	\kappa_f& -\frac{\sigma_{xy}^{(f)}q}{2\pi}\\
	\frac{\sigma_{xy}^{(f)}q}{2\pi} & k_0\frac{|\omega|}{q}+\chi_dq^2
\end{array}
\right)
\eeq
where $\kappa_f$, $k_0$ and $\chi_d$ are non-universal quantities. More precisely, $\kappa_f$ is the compressibility of $f_\ell$, $\sigma^{(f)}_{xy}$ is the Hall conductivity of $f_\ell$, the $k_0|\omega|/q$ term corresponds to Landau damping (for $\omega\ll v_fq$), and the $\chi_d$ is the diamagnetic suspectibility.

Further integrating out the emergent gauge field $a_\ell$ yields the response theory to the physical electromagnetic gauge field $A_\ell$:
\beq
\mc{L}_{\rm response}=\frac{1}{2}A_\ell(k)^T\Pi_c(k)A_\ell(-k)
\eeq
with
\beq \label{eq: Ioffe-Larkin}
\Pi_{c}=\Pi_{b}-2\Pi_s-(\Pi_{b}-\Pi_s)(\Pi_{b}+\Pi_{f})^{-1} (\Pi_{b}-\Pi_s)
\eeq
In the case of no nontrivial coupling in the spectator sector, \ie when $\Pi_s$ can be ignored, as $b_\ell$ forms a QH state and $f_\ell$ forms a Fermi surface for all $C$, the thermodynamic and transport properties obtained from this Ioffe-Larkin-type analysis should be qualitatively the same for all $C$, and the case with $C=1$ precisely corresponds to the standard CFL \cite{HLR}. This expectation still largely holds even when $\Pi_s$ is taken into account.


As an example, let us demonstrate that the gCFL is compressible. The electronic compressibility of $c_\ell$ is $\kappa_c=\Pi_c^{(00)}(\omega=0, \vec q\rightarrow 0)$ . Substituting \eqref{eq: polarization spectator}, \eqref{eq: polarization boson} and \eqref{eq: polarization fermion} into \eqref{eq: Ioffe-Larkin} yields
\beq
\kappa_c=\frac{\kappa_f(\sigma_{xy}^{(b)}-2)^2}{4\pi^2\kappa_f\chi_d+(\sigma_{xy}^{(b)}+\sigma_{xy}^{(f)})^2}
\eeq
Thus, $\kappa_{c}\neq 0$ and the system is compressible as long as $b_\ell$ is in a QH state (with $\sigma_{xy}^{(b)}\neq2$) and $f_\ell$ forms a gapless Fermi surface (with $\kappa_f\neq 0$). Notice the factor $(\sigma_{xy}^{(b)}-2)^2$ in the numerator above is due to $\Pi_s$; if $\Pi_s=0$, this factor becomes simply $(\sigma_{xy}^{b})^2$.

One notable distinction between the gCFL and the usual CFL is in their spectral properties. To understand it systematically, consider its low-energy effective theory after integrating out the bosonic sector:
\beq
\mc{L}_\ell=\mc{L}_{[f_\ell, a_\ell+A_s]}+\frac{\nu}{4\pi}(a_\ell-A_\ell) d(a_\ell-A_\ell)+\mc{L}_{[c_\ell, B_\ell]}
\eeq

To examine the electronic spectral property, we note that an electron $c_\ell$ is a bound state of gapped $b_\ell$ and gapless $f_\ell$. Naively one expects that the single electron is always gapped because $b_\ell$ is gapped. However, because $b_\ell$ is in a QH state with $\sigma_{xy}^{(b_\ell)}=\nu$, we may also be able to make an electron by binding $f_\ell$ with certain monopoles of $a_\ell$. This is indeed the case when $C=1$, where we can make up an electron by binding 2 anti-monopoles of $a_\ell$ and one $f_\ell$ quantum (with or without the presence of the spectator sector captured by $\mc{L}_{[c_\ell, B_\ell]}$). As a result, a single electron has a soft gap \cite{BH93, Kim1994}. Applying the same reasoning to a gCFL with $C\neq 1$, one may attempt to identify a single electron with a bound state of $f_\ell$ and $1/\nu=(C+1)/C$ anti-monopoles of $a_\ell$. But there can only be an integral number of monopoles of $a_\ell$, so a $1/\nu$ anti-monopole does not exist if $C\neq 1$. Therefore, a single electron has a hard gap when $C\neq 1$. However, one can make a local multi-electron bound state by binding $C$ $f_\ell$'s with $C+1$ anti-monopoles of $a_\ell$, which then has a soft gap. {\footnote{Interestingly, this bound state has electric charge $C-2$ and $S^z$-spin $C$, so it can be viewed as a bound state of $C-1$ spin-up electrons and 1 spin-down hole.}} More generally, only $N$ composites of such multi-electron bound states have soft gaps in a gCFL with $C\neq 1$, where $N$ is an integer, {\footnote{Collective modes arising from Fermi surface deformation belong to the case with $N=0$.}} and all other excitations have hard gaps. This is similar to a $Z_C$ OM, which has qualitatively the same thermodynamic and transport properties as a usual Fermi liquid. However, the single electron in a $Z_C$ OM is gapped, and only $NC$-electron bound states are gapless, with $N$ an integer \cite{Nandkishore2012}. 

We note so far our gCFL is defined at filling factors specified in the main text, but actually it can also be properly constructed at filling $\nu=p/q$ in a similar way, with $p$ and $q$ more generic mutually co-prime integers. Such a gCFL still has similar thermodynamic and transport properties as the usual CFL, but only multi-electron bound states have soft gaps, and excitations with other charges have a hard gap. Indeed, such a state with $p$ being an even integer and $q=1$ was recently constructed in Ref. \cite{Zhang2020}.

\section{Spectator sector and the internal gauge flux}

In the main text we mentioned that the valence spin-up electrons and Cooper pairs are gapped spectators to the transition, but they play an important role in that their coupling can fix the internal gauge flux of $a_\ell$ without fine-tuning. In this section we provide more details of the spectator sector and the internal gauge flux. We will focus on the single-layer case and it is straightforward to generalize the discussion to the bilayer case.

Let us first consider the case where the spectators are gapped but there is no nontrivial coupling between them. At the critical point, from the equations of motion of $a_\ell$ and $\alpha_\ell$ we get
\beq \label{eq: EOM app}
\frac{\delta b_a+\delta b_\alpha}{2\pi}+\delta n_f=0,
\quad
-\frac{\delta b_a+\delta b_\alpha}{2\pi}+\delta n_\psi+\frac{C-1}{4\pi}\delta b_\alpha=0
\eeq
with $\delta b_a$ the deviation from the value of $b_a$ at the putative critical point, where $b_a$ is the flux of $a_\ell$. Similarly for $\delta b_\alpha$, $\delta n_f$ and $\delta n_\psi$. By varying the Lagrangian with respect to $A_\ell$ and $A_s$, we can obtain the electric charge and total spin on the $\ell$-th layer. Since these are fixed, we get two more conditions:
\beq \label{eq: fixed charge app}
-\frac{\delta b_a+\delta b_\alpha}{2\pi}=0,
\quad
\delta n_f=0
\eeq
These are all the conditions we have, and it is easy to see that with these conditions it is impossible to ensure that all of $\delta b_a$, $\delta b_\alpha$, $\delta n_f$ and $\delta n_\psi$ vanish. In particular, although $n_f$ is fixed, $b_a$, $b_\alpha$ and $n_\psi$ can still change without violating the above conditions. Once $b_a\neq 0$, the critical point is expected to be unstable to form a QH state, rendering the single-layer FL-gCFL indirect.

More generally, without nontrivial coupling in the spectator sector, the current setup is essentially the same as the spinless setup in Ref. \cite{Barkeshli2012}. Then we need to fix 3 quantities: the densities of the two partons as well as the internal flux density. However, there are only 2 conditions: gauge neutral condition \eqref{eq: EOM app} and fixed electric charge condition \eqref{eq: fixed charge app}. So in this case it is impossible to fix the 3 quantities simultaneously, and the transition requires tuning two parameters. This type of fine-tuning is expected to be a general feature of transitions associated with multiple species of gapless partons coupled to a U(1) gauge field, unless the system has enough symmetries to fix all these quantities.

In passing, it is worth mentioning why the internal gauge flux is fixed in the gCFL phase even without the spectator sector, which is necessary for the spinless version of this phase to be stable. The reason is that in the gCFL phase the bosonic partons are gapped, so their density cannot be changed under weak perturbations. Then there are only 2 quantities to fix, the density of the fermionic parton and the internal gauge flux density, and it is possible to fix them just by the two conditions above. Similarly, the transition within the bosonic sector also only involves one gapless parton and one dynamical U(1) gauge field, $\alpha_\ell$, (recall that $A_\ell-a_\ell$ is regarded as a static gauge field when discussing the transition within the bosonic sector), and these can be fixed by the gauge neutral condition and fixed electric charge condition.

Now let us see how a coupling $\mc{L}_{[c_\ell, B_\ell]}=-\frac{2}{2\pi}A_\ell d(A_\ell-a_\ell)$ helps us fix all the 3 quantities. Intuitively, in this case, the flux of $a_\ell$ now carries charge under $A_\ell$, so it is fixed since the charge under $A_\ell$ is fixed. More formally, we see that the gauge neutral condition resulted from the equations of motion of $a_\ell$ and $\alpha_\ell$ are still given by \eqref{eq: EOM app}, but the fixed charge condition becomes
\beq
-\frac{\delta b_a+\delta b_\alpha}{2\pi}+\frac{\delta b_a}{\pi}=0,
\quad
\delta n_f=0
\eeq
Now these equations ensure that $\delta b_a=\delta b_\alpha=\delta n_f=\delta n_\psi=0$. That is, all these quantities are fixed to their presumed values at the critical point, and this is a necessary condition for the transition to be accessible by tuning a single parameter.

Finally, we elaborate on how the interaction $\mc{L}_{[c_\ell, B_\ell]}=-\frac{2}{2\pi}A_\ell d(A_\ell-a_\ell)$ can emerge. For this purpose, let us momentarily assume that the conduction electrons and the valance electrons are actually two decoupled systems, and we denote their annihilation operators by $c_{\ell c}$ and $c_{\ell v}$, and the U(1) gauge fields they coupled to by $A_{\ell c}$ and $A_{\ell v}$, respectively. Then we perform the parton construction $c_{\ell c}=b_{\ell c}f_{\ell c}$ and $c_{\ell v}=b_{\ell v}f_{\ell v}$, which implies emergent gauge fields $a_{\ell c}$ and $a_{\ell v}$ for conduction and vallance partons, respectively. Then the low-energy effective theory takes the form
\beq
\mc{L}_\ell=\mc{L}_{[b_{\ell c}, A_{\ell c}-a_{\ell c}]}+\mc{L}_{[f_{\ell c}, a_{\ell c}+A_s]}+\mc{L}_{[b_{\ell v}, A_{\ell v}-a_{\ell v}]}+\mc{L}_{[f_{\ell v}, a_{\ell v}+A_s]}+\mc{L}_{[B_\ell, \tilde A]}+\cdots
\eeq
where $\cdots$ represents the coupling between the various sectors, and at this point we denote the gauge field coupled to $B_\ell$ by $\tilde A$. Now we can attach a vortex of $B_\ell$ to $b_{\ell v}$, and attach a vortex of $b_{\ell v}$ to $B_\ell$. According to Ref. \cite{Senthil2012}, this induces an interaction that takes the following form at low energies:
\beq \label{eq: BSPT interaction}
\mc{L}_{[b_{\ell v}, B_\ell]}=-\frac{1}{2\pi}\tilde Ad(A_{\ell v}-a_{\ell v})
\eeq
Finally, we switch on weak hybridization between $b_{\ell v}$ and $b_{\ell c}$, between $f_{\ell v}$ and $f_{\ell c}$, and between $B_\ell$ and the bound state of the physical spin-up and spin-down electrons. As long as this hybridization strength is weaker than the gap of spectator sectors, $c_{\ell v}$ and $B_\ell$, its main effect is to identify $a_{\ell c}=a_{\ell v}=a_\ell$, $A_{\ell c}=A_{\ell v}=A_\ell$ and $\tilde A=2\tilde A_\ell$. In particular, the interaction \eqref{eq: BSPT interaction} simply becomes
\beq
\mc{L}_{[c_\ell, B_\ell]}=-\frac{2}{2\pi}A_\ell d(A_\ell-a_\ell)
\eeq
Notice in terms of the electronic degrees of freedom, this hybridization, albeit weak, requires reasonably strong interactions.

\section{Interlayer pairing instability}

Here we generalize the results of Refs. \cite{Zou2020, Bonesteel1996, Sodemann2016, Isobe2016} to show that two layers of gCFL are unstable towards interlayer pairing between $f_1$ and $f_2$ in the presence of a repulsive long-range interaction potential of the form $V(r)=V_0/r^{1+\epsilon}$, if $0\leqslant\epsilon<1$. 

Similar to Ref. \cite{Sodemann2016}, we can write the long-range interaction as
\beq
\begin{split}
	\mc{L}_{\rm LR}
	=&\sum_{\ell_1 \ell_2}(\nabla\times a_{\ell_1t})[\vec r_i]\frac{V_0}{\left[(1-\delta_{\ell_1\ell_2})d^2+(\vec r_i-\vec r_j)^2\right]^{\frac{1+\epsilon}{2}}}(\nabla\times a_{\ell_2t})[\vec r_j]\\
	=&(\nabla\times a_{+t})[\vec r_i]f_+(\vec r_i-\vec r_j, d)(\nabla\times a_{+t})[\vec r_j]+(\nabla\times a_{-t})[\vec r_i]f_-(\vec r_i-\vec r_j, d)(\nabla\times a_{-t})[\vec r_j]
\end{split}
\eeq
with
\beq
\begin{split}
	&f_+(\vec r, d)=2V_0 \left[ \frac{1}{r^{1+\epsilon}}+\frac{1}{\left(d^2+r^2\right)^{(1+\epsilon)/2}}\right]\stackrel{r\gg d}{\longrightarrow} \frac{4V_0}{r^{1+\epsilon}}\\
	&f_-(\vec r, d)=2V_0 \left[\frac{1}{r^{1+\epsilon}}-\frac{1}{\left(d^2+r^2\right)^{(1+\epsilon)/2}}\right]\stackrel{r\gg d}{\longrightarrow}V_0(1+\epsilon)\frac{d^2}{r^{3+\epsilon}}.
\end{split}
\eeq
As in the main text, $a_{\ell t}$ represents the transverse spatial component of $a_{\ell}$. 

Transforming to momentum space, in the large-distance limit this interaction contributes ``kinetic" terms to $a_\pm$:
\beq
S_{\rm LR}[a_{+t}]=\int_\omega\int_{\vec k}\frac{|k|^{1+\epsilon}}{2g_+}|a_{+t}(\vec k, \omega)|^2,
\quad
S_{\rm LR}[a_{-t}]=\int_\omega\int_{\vec k}\frac{|k|^{3+\epsilon}}{2g_-}|a_{-t}(\vec k, \omega)|^2,
\eeq
where $g_+\sim 1/V_0$ and $g_-\sim 1/(V_0 d^2(1+\epsilon))$. Notice that local interactions will automatically generate kinetic terms for $a_\pm$ of the form
\beq
S_{\rm local}[a_{\pm t}]=\int_\omega\int_{\vec k}\frac{k^2}{2g_\pm'}|a_{\pm t}(\vec k, \omega)|^2.
\eeq
At low energies $S_{\rm local}[a_{-t}]$ dominates over $S_{\rm LR}[a_{-t}]$ for all $\epsilon\geqslant 0$. On the other hand, when $0\leqslant \epsilon<1$, at low energies $S_{\rm LR}[a_{+t}]$ dominates over $S_{\rm local}[a_{+t}]$, while the former is equally or less relevant than the latter when $\epsilon\geqslant 1$. Therefore, for the case of primary interest to us, where $0\leqslant\epsilon<1$, we will only keep $S_{\rm LR}[a_{+t}]$ and $S_{\rm local}[a_{-t}]$ in the low-energy regime:
\beq
S[a_{+t}, a_{-t}]=\int_\omega\int_{\vec k}\left[\frac{|k|^{1+\epsilon_+}}{2g_+}|a_{+t}(\vec k, \omega)|^2+\frac{k^{1+\epsilon_-}}{2g_-}|a_{-t}(\vec k, \omega)|^2\right].
\eeq
with $\epsilon_+=\epsilon$ and $\epsilon_-=1$. Notice we have dropped all primes in the effective gauge couplings and simply write them as $g_\pm$.

The effect between the coupling of the gauge fields and fermions can be captured by the following theory \cite{Nayak1994, Lee2008, Lee2009, Metlitski2010, Mross2010, Metlitski2014, Zou2020}:
\beq
S=S_f+S[a_{+t}, a_{-t}],
\eeq
where $S_f$ is given by $S_f=\int d^2xd\tau\mc{L}_f$, with
\beq
\mc{L}_f=\sum_{p=\pm,\ell=1,2}f^\dagger_{\ell p}\left[\partial_\tau + v_F\left(-ip\partial_x -\partial_y^2\right)\right]f_{\ell p}- v_F \sum_{p=\pm}\lambda_p\left[(a_++a_-)f_{1p}^\dag f_{1p}+(a_+-a_-)f_{2p}^\dag f_{2p}\right],
\eeq
where $f_{\ell p}$ is the low-energy mode of the fermions in the $\ell$th layer, near antipodal patches on the Fermi surface labeled by $p~(=\pm)$.

For the above theory, $a_+$ tends to suppress pairing while $a_-$ tends to promote interlayer pairing \cite{Zou2020, Bonesteel1996, Zou2016, Sodemann2016, Isobe2016}. From the above action we can see that when $\epsilon<1$, the coupling between $a_-$ and the fermions is more RG relevant than the coupling between $a_+$ and the fermions, so we expect $a_-$ will win over $a_+$ in the competition, and the Fermi surface becomes unstable to interlayer pairing. 

To see this more formally, define dimensionless gauge couplings $\tilde g_\pm\equiv \frac{g_\pm v_F}{4\pi^2\Lambda^{\epsilon_\pm}}$ as in Ref. \cite{Zou2020}, where $\Lambda$ is a cutoff scale of this effective theory. Denote the dimensionless interlayer 4-fermion interaction by $\widetilde V$. We will study the RG flow of these couplings in an expansion in terms of small $\epsilon_+$ and $\epsilon_-$, with the assumption that the physics at $\epsilon_+=\epsilon$ and $\epsilon_-=1$ is qualitatively the same as that for small $\epsilon_\pm$. Generalizing the results in Ref. \cite{Zou2020}, we get the beta functions of these couplings:{\footnote{We note that the corresponding beta functions in Ref. \cite{Sodemann2016} did not include some of these terms.}}
\beq \label{eq: beta function full}
\begin{split}
	&\beta(\tilde g_+)=\left(\frac{\epsilon_+}{2}-(\tilde g_++\tilde g_-)\right)\tilde g_+\\
	&\beta(\tilde g_-)=\left(\frac{\epsilon_-}{2}-(\tilde g_++\tilde g_-)\right)\tilde g_-\\
	&\beta(\widetilde V)=\tilde g_+-\tilde g_--\widetilde V^2
\end{split}
\eeq
The first two equations have a single attractive fixed point at $(\tilde g_+, \tilde g_-)=(0, 1/2)$ when $0\leqslant\epsilon<1$. Notice that $\tilde g_+\rightarrow 0$ under RG agrees with our previous expectation that this coupling is irrelevant. Substituting this result into the last equation, we see that $\widetilde V$ flows to $-\infty$, which signals an interlayer pairing.{\footnote{We note that the next-leading-order correction to the third beta function is of the form $\tilde g_\pm\cdot\widetilde V$, resulting from vertex correction and the flow of Fermi velocity \cite{Metlitski2014, Zou2020}. For small $\epsilon_\pm$, such corrections cannot change the results that the Fermi surfaces are unstable to interlayer pairing.}} Notice that the precise channel in which pairing occurs depends on non-universal details \cite{Metlitski2014, Sodemann2016, Isobe2016}.

\section{RG irrelevance of interlayer couplings at the metal-insulator quantum critical point}

In this section, we provide additional details underlying the effects of a variety of interlayer couplings at the metal-insulator transition in the QH bilayer by applying the method in Ref. \cite{Zou2016}: enumerating all possible interlayer couplings and analyzing their effects at the critical point. We will see that a sufficient condition for the layer decoupling is that all gauge invariant operators in the $b_\ell$-sector have a scaling dimension larger than $3/2$, which is satisfied when $C\gg 1$ \cite{Chen1993}.

Below we enumerate and analyze some of the most relevant interlayer couplings.

\begin{enumerate}
	
	\item $\left(\mc{O}_{b_1}\cdot f_2^\dag f_2\right)+(1\leftrightarrow 2)$, where the first (second) term is a coupling between a gauge invariant operator in the $b_1$ ($b_2$)-sector and a fermion bilinear in the $f_2$ ($f_1$)-sector. For our purpose, it is sufficient to focus on the first term. Let us integrate out $f_2$ and examine the effect of this term on the $b_1$-sector. Due to the first term, integrating out $f_2$ generates an effective action of the form $\int_{\omega, \vec k}|\mc{O}_{b_1}(\vec k, \omega)|^2$ in the critical regime of $b_1$ with $\omega\sim k$. Clearly this effective action is irrelevant when the scaling dimension of $\mc{O}_{b_1}$ is larger than $3/2$, which is satisfied when $C\gg 1$. 
	
	Next we integrate out $b_1$ and examine the effect of this term on the $(f_2, a_2)$-sector. This generates an effective action of the form,
	\beq
	\int_{\omega, \vec k} |\omega^2+k^2|^{\Delta_{\mc{O}_{b_1}}-\frac{3}{2}}|f_2^\dag f_2(\vec k, \omega)|^2
	\sim
	\int_{\omega,\vec k} |k_y|^{2\Delta_{\mc{O}_{b_1}}-3}|f_2^\dag f_2(\vec k, \omega)|^2,
	\eeq
	where $\Delta_{\mc{O}_{b_1}}$ is the scaling dimension of $\mc{O}_{b_1}$, and we have used the scaling relation $\omega\sim k_y^{z_a}\sim k_x^{z_a/2}$, with $k_x$ the momentum parallel to the Fermi velocity and $k_y$ the momentum perpendicular to the Fermi velocity \cite{Lee2008, Lee2009, Metlitski2010, Mross2010}. The gauge fields contribute \cite{Lee2008, Lee2009, Metlitski2010, Mross2010}
	\beq
	\sim\int_{\omega,\vec k}\frac{1}{|k_y|^{z_a-1}}|f_2^\dag f_2(\vec k, \omega)|^2
	\eeq
	So, compared to this gauge-field contribution, the interlayer coupling is irrelevant if $\Delta_{\mc{O}_{b_1}}>2-\frac{z_a}{2}=1$, where we have used $z_a=2$. This is again satisfied when $C\gg 1$.
	
	\item $\left(\mc{O}_{b_1}\cdot\nabla\times a_2\right)+(1\leftrightarrow 2)$. Since the gauge field is Landau damped, this coupling is irrelevant to the bosonic sector \cite{Senthil2008}. On the other hand, integrating out $b_1$ generates an effective action for $a_2$ of the form $k^{2\Delta_{\mc{O}}-1}|a_2(\vec k, \omega)|^2$, where $\Delta_{\mc{O}}$ is the scaling dimension of $\mc{O}$. This term is  RG irrelevant compared to the gauge field effective action due to the critical boson when $\Delta_{\mc{O}}>1$, which is satisfied when $C\gg 1$. So this coupling is also irrelevant.
	
	\item $(\nabla\times a_1)(\nabla\times a_2)$. Because of the presence of the gauge field effective action due to the critical boson, this coupling is irrelevant.
	
	\item $(f_1^\dag f_1\cdot\nabla\times a_2)+(1\leftrightarrow 2)$. This coupling has one more derivative compered to the minimal coupling between $a_1$ and $f_1$, which has a finite density. So this coupling is irrelevant.
	
	\item Interlayer electron tunneling, which contributes the following low-energy effective action:
	\beq
	\delta S_{\rm tunneling}=-\int d\omega dk_\perp d\theta \lambda(\theta)\left(c_1^\dag(k_\perp, \omega, \theta) c_2(k_\perp, \omega, \theta) +\hc\right)
	\eeq
	where $k_\perp$ is the distance away from the Fermi surface in the momentum space, $\theta$ parameterizes the position on the Fermi surface, $\lambda$ is a $\theta$-dependent interlayer electron tunneling strength, and $c_{1, 2}(k_\perp, \omega, \theta)$ represents the low-energy electron operators near the Fermi surface. Consider the scaling transformation $k_\perp\rightarrow k'_\perp=sk_\perp$, $\omega\rightarrow\omega'=s\omega$ and $\theta\rightarrow\theta'=\theta$. Note that the leading singular contribution to the electron spectral function at the critical point is given by \cite{Senthil2008}:
	\beq
	\mc{A}_c(k_\perp, \omega, \theta)=\frac{\omega^\eta}{\ln\frac{\Lambda_\omega}{\omega}}\cdot F\left(\frac{\omega\ln\frac{\Lambda_\omega}{\omega}}{vk_\perp}\right)
	\eeq
	with the universal function $F(x)=\left(1-\frac{1}{x}\right)^\eta\theta(x-1)$, where $\eta$ is the anomalous dimension of $b_\ell$ in the transition described by 
	\eqref{eq: bosonic transition supp}, $\Lambda_\omega$ and $v$ are non-universal constants. This spectral function implies that under the above scaling transformation, $c_{\ell}(k_\perp, \omega, \theta)\rightarrow c_{\ell}'(k'_\perp, \omega', \theta')=s^{\frac{\eta}{2}-1}c_\ell(k, \omega, \theta)$ and $\lambda(\theta)\rightarrow\lambda'(\theta')=s^{-\eta}\lambda(\theta)$. Because $\eta>0$, interlayer electron tunneling is also irrelevant at the transition.{\footnote{When $b_\ell$ is gapped, the electron $c_\ell$ is even more incoherent, and interlayer electron tunneling will still be irrelevant. When $b_\ell$ is condensed, each layer is in a Fermi liquid phase, so interlayer electron tunneling is relevant. Therefore, interlayer electron tunneling is a dangerously irrelevant perturbation.}}
	
\end{enumerate}

\section{$K$-matrix of interlayer paired state in a channel with $n\neq\pm 1$}

In this section we derive the $K$-matrix for interlayer paired states in a channel with $n\neq 1$, and all we need is to modify the effective theory for the $d_\ell$ fermions. In these channels, it is not immediately obvious how $a_-$ should couple with other gauge fields, because the arguments in Ref. \cite{Sodemann2016}, involving the layer exchange symmetry and the zero modes in the vortex cores of these paired states, do not directly apply here. So the simplest and systematic way to proceed may be to apply the trick in Ref. \cite{Zou2018} to build up the theory for a state with $n\neq\pm 1$ from the states with $n=\pm 1$. The key observation behind this approach is that switching on the hybridization between fermions in a stack of states with $n=\pm 1$ can produce a state with any $n$ \cite{Kitaev2006}.

For example, for the $n=0$ state, we can view it as a $n=1$ state together with a $n=-1$ state, where the fermions $d_\ell$ in the $n=1$ state can hybridize with those in the $n=-1$ state. Before considering the hybridization between these $d_\ell$ fermions,
\beq
\mc{L}_{[d_\ell, a_-]}=-\frac{4}{4\pi}\gamma_1d\gamma_1+\frac{1}{\pi}a_-d\gamma_1+\frac{4}{4\pi}\gamma_2d\gamma_2+\frac{1}{\pi}a_-d\gamma_2
\eeq
where the first two terms represent the $n=1$ state, and the last two terms represent the $n=-1$ state. Notice this theory actually has two dynamical $Z_2$ gauge field, where the charge-1 excitation of $\gamma_1$ is identified with the $\pi$-flux of one $Z_2$ gauge field, and the charge-1 excitation of $\gamma_2$ is identified with the $\pi$-flux of the other $Z_2$ gauge field. The hybridization between the fermions in the $n=1$ and $n=-1$ states confines these two types of $\pi$-flux together, which can be formally implemented by imposing the constraint $q_{\gamma_1}=q_{\gamma_2}\ ({\rm mod\ } 2)$. As a sanity check, let us introduce $\gamma_1'=\gamma_1+\gamma_2$ and $\gamma_2'=\gamma_1-\gamma_2$, whose charges can independently take any integer. In terms of $\gamma_1'$ and $\gamma_2'$, the above theory with $a_-$ switched off is $-\frac{1}{\pi}\gamma_1'd\gamma_2'$, which, as expected, is precisely the standard $K$-matrix description of an $n=0$ superconductor coupled to a dynamical $Z_2$ gauge field.

Combining $\phi$ and $d_\ell$, we get the effective theory of $f_\ell$:
\beq \label{eq: channel n=0}
\mc{L}_{[f, a+A_s]}=\frac{1}{\pi}(a_++A_s)d\beta
-\frac{4}{4\pi}(\gamma_1 d\gamma_1-\gamma_2d\gamma_2)+\frac{1}{\pi}a_-d(\gamma_1+\gamma_2)
\eeq
with a constraint $q_\beta=q_{\gamma_1}=q_{\gamma_2}\ ({\rm mod\ } 2)$. To eliminate this constraint, we can introduce $\tilde\beta$, $\tilde\gamma_1$ and $\tilde\gamma_2$ as 
\beq
\left(
\begin{array}{c}
	\tilde\beta\\
	\tilde\gamma_1\\
	\tilde\gamma_2
\end{array}
\right)
=
\left(
\begin{array}{ccc}
	1 & -1 & 1\\
	1 & 1 & -1\\
	-1 & 1 & 1
\end{array}
\right)
\cdot
\left(
\begin{array}{c}
	\beta\\
	\gamma_1\\
	\gamma_2
\end{array}
\right).
\eeq
The charges of $\tilde\beta$, $\tilde\gamma_1$ and $\tilde\gamma_2$ can independently take all integers.

As in the main text, introducing gauge fields $\beta_\ell^i$ to rewrite the effective theory of $b_\ell$ described by (\ref{eq: boson QH effective supp}), and combining the result with \eqref{eq: channel n=0} and $\mc{L}_{[c_\ell, B_\ell]}=-\frac{2}{2\pi}A_\ell d(A_\ell-a_\ell)$, we get the effective theory of $c_\ell$:
\beq
\mc{L}=\frac{K_{IJ}}{4\pi}\tilde a_Id\tilde a_J+\frac{t_{1I}}{2\pi}A_1d\tilde a_I+\frac{t_{2I}}{2\pi}A_2d\tilde a_I+\frac{t_{sI}}{2\pi}A_sd\tilde a_I
\eeq
where $\tilde a_I=(\beta_1^1, \beta_1^2, \cdots, \beta_1^{C+1}, \beta_2^1, \beta_2^2, \cdots, \beta_2^{C+1}, \tilde\beta, \tilde\gamma_1, \tilde\gamma_2)^T$, $t_1=(0, 0, \cdots, 0, 1, 1, 1)^T$, $t_2=(0, 0, \cdots, 0, -1)^T$, and
\beq
\begin{split}
	&K=
	\left(
	\begin{array}{ccc}
		K_1 & 0 & K_2 \\
		0 & K_1 & K_3 \\
		K_2^T & K_3^T & K_4
	\end{array}
	\right)\\
	&t_1=(-2, -2, \cdots, -2, 0, 0, \cdots, 0, 1, 1, 1)^T\\
	&t_2=(0, 0, \cdots, 0, -2, -2, \cdots, -2, 0, 0, -1)^T\\
	&t_s=(0, 0, \cdots, 0, 1, 1, 0)
\end{split}
\eeq
with $K_1$ a $(C+1)\times(C+1)$ matrix whose diagonal entries all vanish and other entries are all 1, $K_2$ a $(C+1)\times 3$ matrix whose all entries are 1, $K_3$ a $(C+1)\times 3$ matrix whose first two columns vanish and all entries in the third column are $-1$, and
\beq
K_4=
\left(
\begin{array}{ccc}
	1 & 0 & 1 \\
	0 & -1 & -1 \\
	1 & -1 & 0
\end{array}
\right).
\eeq

The $K$-matrix may be simplified by introducing a new set of gauge field such that $\tilde a_I=W_{IJ}a'_J$, where $W\in SL(2, \mathbb{Z})$. In terms of $a'$, the new $K$-matrix is $K'=W^TKW$, and the new $t_{1, 2}$ vector is $t'_{1,2,s}=W^Tt_{1,2,s}$ \cite{Wen2004Book}. For example, when $C=1$, take
\beq
W=
\left(
\begin{array}{ccccccc}
	1 & 0 & 1 & 0 & 0 & 0 & 0 \\
	0 & 1 & 1 & 0 & 0 & 0 & 0 \\
	0 & 0 & 1 & 0 & 0 & 0 & 0 \\
	-1 & -1 & -3 & 1 & 1 & 0 & 0 \\
	-1 & -1 & -3 & 0 & 1 & 0 & 0 \\
	1 & 1 & 1 & -1 & 0 & 1 & -1 \\
	0 & 0 & 1 & 1 & 0 & 0 & 1 \\
\end{array}
\right)
\eeq
We get a simplified $K'=W^TKW={\rm diag}(\sigma_x, -4, -1, 1, -1, 1)$ where $\sigma_x$ is the standard Pauli matrix. That is, the resulting state has a U$(1)_4$ topological order. 

In passing, we note that for the case with $C=1$ and $n=-1$, take
\beq
W=
\left(
\begin{array}{cccccc}
	1 & 0 & 0 & -1 & 1 & 0 \\
	0 & 1 & 0 & -1 & 1 & 0 \\
	0 & 0 & 1 & 1 & -1 & 0 \\
	0 & 0 & 0 & 1 & 0 & 0 \\
	0 & 0 & 0 & 1 & -1 & 0 \\
	0 & 0 & -1 & -1 & 0 & 1 \\
\end{array}
\right)
\eeq
we can convert the $K$-matrix in the main text into
\beq
K'=W^TKW=
\left(
\begin{array}{cccccc}
	0 & 1 & 0 & 0 & 0 & 0 \\
	1 & 0 & 0 & 0 & 0 & 0 \\
	0 & 0 & -1 & 0 & 0 & 0 \\
	0 & 0 & 0 & 0 & 0 & 0 \\
	0 & 0 & 0 & 0 & -1 & 0 \\
	0 & 0 & 0 & 0 & 0 & 1 \\
\end{array}
\right)
\eeq
together with $t_1'=(-2, -2, 0, 5, -5, 0)^T$, $t_2'=(0, 0, -3, -5, 2, 1)^T$, and $t_s'=(0, 0, -1, -1, 0, 1)^T$, which represents a state that spontaneously breaks a U(1) symmetry generated by $5(Q_1-Q_2)-S^z$, where $Q_{1,2}$ is the electric charge of the two layers, respectively.

This approach can be similarly applied to obtain the $K$-matrix description of states for any $n$, and one only has to change $\mc{L}_{[d_\ell, a_-]}$ appropriately. For instance, if $n=n_0>0$ ($n=n_0<0$), one can start with $|n_0|$ of $n=1$ ($n=-1$) states and switch on the hybridization among fermions in different $n=1$ ($n=-1$) components \cite{Kitaev2006}.

\bibliography{supplementary.bib}

\begin{thebibliography}{49}%
\makeatletter
\providecommand \@ifxundefined [1]{%
 \@ifx{#1\undefined}
}%
\providecommand \@ifnum [1]{%
 \ifnum #1\expandafter \@firstoftwo
 \else \expandafter \@secondoftwo
 \fi
}%
\providecommand \@ifx [1]{%
 \ifx #1\expandafter \@firstoftwo
 \else \expandafter \@secondoftwo
 \fi
}%
\providecommand \natexlab [1]{#1}%
\providecommand \enquote  [1]{``#1''}%
\providecommand \bibnamefont  [1]{#1}%
\providecommand \bibfnamefont [1]{#1}%
\providecommand \citenamefont [1]{#1}%
\providecommand \href@noop [0]{\@secondoftwo}%
\providecommand \href [0]{\begingroup \@sanitize@url \@href}%
\providecommand \@href[1]{\@@startlink{#1}\@@href}%
\providecommand \@@href[1]{\endgroup#1\@@endlink}%
\providecommand \@sanitize@url [0]{\catcode `\\12\catcode `\$12\catcode
  `\&12\catcode `\#12\catcode `\^12\catcode `\_12\catcode `\%12\relax}%
\providecommand \@@startlink[1]{}%
\providecommand \@@endlink[0]{}%
\providecommand \url  [0]{\begingroup\@sanitize@url \@url }%
\providecommand \@url [1]{\endgroup\@href {#1}{\urlprefix }}%
\providecommand \urlprefix  [0]{URL }%
\providecommand \Eprint [0]{\href }%
\providecommand \doibase [0]{http://dx.doi.org/}%
\providecommand \selectlanguage [0]{\@gobble}%
\providecommand \bibinfo  [0]{\@secondoftwo}%
\providecommand \bibfield  [0]{\@secondoftwo}%
\providecommand \translation [1]{[#1]}%
\providecommand \BibitemOpen [0]{}%
\providecommand \bibitemStop [0]{}%
\providecommand \bibitemNoStop [0]{.\EOS\space}%
\providecommand \EOS [0]{\spacefactor3000\relax}%
\providecommand \BibitemShut  [1]{\csname bibitem#1\endcsname}%
\let\auto@bib@innerbib\@empty
\bibitem [{\citenamefont {Senthil}\ \emph {et~al.}(2004)\citenamefont
  {Senthil}, \citenamefont {Vojta},\ and\ \citenamefont {Sachdev}}]{TSFLstar}%
  \BibitemOpen
  \bibfield  {author} {\bibinfo {author} {\bibfnamefont {T.}~\bibnamefont
  {Senthil}}, \bibinfo {author} {\bibfnamefont {Matthias}\ \bibnamefont
  {Vojta}}, \ and\ \bibinfo {author} {\bibfnamefont {Subir}\ \bibnamefont
  {Sachdev}},\ }\bibfield  {title} {\enquote {\bibinfo {title} {Weak magnetism
  and non-fermi liquids near heavy-fermion critical points},}\ }\href
  {http://dx.doi.org/10.1103/PhysRevB.69.035111} {\bibfield  {journal}
  {\bibinfo  {journal} {Physical Review B}\ }\textbf {\bibinfo {volume} {69}}
  (\bibinfo {year} {2004})}\BibitemShut {NoStop}%
\bibitem [{\citenamefont {{Senthil}}(2008{\natexlab{a}})}]{Senthil2008}%
  \BibitemOpen
  \bibfield  {author} {\bibinfo {author} {\bibfnamefont {T.}~\bibnamefont
  {{Senthil}}},\ }\bibfield  {title} {\enquote {\bibinfo {title} {{Theory of a
  continuous Mott transition in two dimensions}},}\ }\href {\doibase
  10.1103/PhysRevB.78.045109} {\bibfield  {journal} {\bibinfo  {journal}
  {\prb}\ }\textbf {\bibinfo {volume} {78}},\ \bibinfo {eid} {045109} (\bibinfo
  {year} {2008}{\natexlab{a}})},\ \Eprint {http://arxiv.org/abs/0804.1555}
  {arXiv:0804.1555 [cond-mat.str-el]} \BibitemShut {NoStop}%
\bibitem [{\citenamefont {Kanoda}\ and\ \citenamefont {Kato}(2011)}]{KKrev}%
  \BibitemOpen
  \bibfield  {author} {\bibinfo {author} {\bibfnamefont {Kazushi}\ \bibnamefont
  {Kanoda}}\ and\ \bibinfo {author} {\bibfnamefont {Reizo}\ \bibnamefont
  {Kato}},\ }\bibfield  {title} {\enquote {\bibinfo {title} {Mott physics in
  organic conductors with triangular lattices},}\ }\href {\doibase
  10.1146/annurev-conmatphys-062910-140521} {\bibfield  {journal} {\bibinfo
  {journal} {Annual Review of Condensed Matter Physics}\ }\textbf {\bibinfo
  {volume} {2}},\ \bibinfo {pages} {167--188} (\bibinfo {year} {2011})},\
  \Eprint
  {http://arxiv.org/abs/https://doi.org/10.1146/annurev-conmatphys-062910-140521}
  {https://doi.org/10.1146/annurev-conmatphys-062910-140521} \BibitemShut
  {NoStop}%
\bibitem [{\citenamefont {Furukawa}\ \emph {et~al.}(2015)\citenamefont
  {Furukawa}, \citenamefont {Miyagawa}, \citenamefont {Taniguchi},
  \citenamefont {Kato},\ and\ \citenamefont {Kanoda}}]{KK15}%
  \BibitemOpen
  \bibfield  {author} {\bibinfo {author} {\bibfnamefont {Tetsuya}\ \bibnamefont
  {Furukawa}}, \bibinfo {author} {\bibfnamefont {Kazuya}\ \bibnamefont
  {Miyagawa}}, \bibinfo {author} {\bibfnamefont {Hiromi}\ \bibnamefont
  {Taniguchi}}, \bibinfo {author} {\bibfnamefont {Reizo}\ \bibnamefont {Kato}},
  \ and\ \bibinfo {author} {\bibfnamefont {Kazushi}\ \bibnamefont {Kanoda}},\
  }\bibfield  {title} {\enquote {\bibinfo {title} {Quantum criticality of mott
  transition in organic materials},}\ }\href {\doibase 10.1038/nphys3235}
  {\bibfield  {journal} {\bibinfo  {journal} {Nature Physics}\ }\textbf
  {\bibinfo {volume} {11}},\ \bibinfo {pages} {221--224} (\bibinfo {year}
  {2015})}\BibitemShut {NoStop}%
\bibitem [{\citenamefont {{Zou}}\ and\ \citenamefont
  {{Chowdhury}}(2020)}]{Zou2020}%
  \BibitemOpen
  \bibfield  {author} {\bibinfo {author} {\bibfnamefont {Liujun}\ \bibnamefont
  {{Zou}}}\ and\ \bibinfo {author} {\bibfnamefont {Debanjan}\ \bibnamefont
  {{Chowdhury}}},\ }\bibfield  {title} {\enquote {\bibinfo {title} {{Deconfined
  metallic quantum criticality: A U(2) gauge-theoretic approach}},}\ }\href
  {\doibase 10.1103/PhysRevResearch.2.023344} {\bibfield  {journal} {\bibinfo
  {journal} {Physical Review Research}\ }\textbf {\bibinfo {volume} {2}},\
  \bibinfo {eid} {023344} (\bibinfo {year} {2020})},\ \Eprint
  {http://arxiv.org/abs/2002.02972} {arXiv:2002.02972 [cond-mat.str-el]}
  \BibitemShut {NoStop}%
\bibitem [{\citenamefont {Stewart}(2001)}]{Stewart}%
  \BibitemOpen
  \bibfield  {author} {\bibinfo {author} {\bibfnamefont {G.~R.}\ \bibnamefont
  {Stewart}},\ }\bibfield  {title} {\enquote {\bibinfo {title}
  {Non-fermi-liquid behavior in $d$- and $f$-electron metals},}\ }\href
  {\doibase 10.1103/RevModPhys.73.797} {\bibfield  {journal} {\bibinfo
  {journal} {Rev. Mod. Phys.}\ }\textbf {\bibinfo {volume} {73}},\ \bibinfo
  {pages} {797--855} (\bibinfo {year} {2001})}\BibitemShut {NoStop}%
\bibitem [{\citenamefont {Schr{\"o}der}\ \emph {et~al.}(2000)\citenamefont
  {Schr{\"o}der}, \citenamefont {Aeppli}, \citenamefont {Coldea}, \citenamefont
  {Adams}, \citenamefont {Stockert}, \citenamefont {L{\"o}hneysen},
  \citenamefont {Bucher}, \citenamefont {Ramazashvili},\ and\ \citenamefont
  {Coleman}}]{Coleman00}%
  \BibitemOpen
  \bibfield  {author} {\bibinfo {author} {\bibfnamefont {A.}~\bibnamefont
  {Schr{\"o}der}}, \bibinfo {author} {\bibfnamefont {G.}~\bibnamefont
  {Aeppli}}, \bibinfo {author} {\bibfnamefont {R.}~\bibnamefont {Coldea}},
  \bibinfo {author} {\bibfnamefont {M.}~\bibnamefont {Adams}}, \bibinfo
  {author} {\bibfnamefont {O.}~\bibnamefont {Stockert}}, \bibinfo {author}
  {\bibfnamefont {H.v.}\ \bibnamefont {L{\"o}hneysen}}, \bibinfo {author}
  {\bibfnamefont {E.}~\bibnamefont {Bucher}}, \bibinfo {author} {\bibfnamefont
  {R.}~\bibnamefont {Ramazashvili}}, \ and\ \bibinfo {author} {\bibfnamefont
  {P.}~\bibnamefont {Coleman}},\ }\bibfield  {title} {\enquote {\bibinfo
  {title} {Onset of antiferromagnetism in heavy-fermion metals},}\ }\href
  {\doibase 10.1038/35030039} {\bibfield  {journal} {\bibinfo  {journal}
  {Nature}\ }\textbf {\bibinfo {volume} {407}},\ \bibinfo {pages} {351--355}
  (\bibinfo {year} {2000})}\BibitemShut {NoStop}%
\bibitem [{\citenamefont {Shishido}\ \emph {et~al.}(2005)\citenamefont
  {Shishido}, \citenamefont {Settai}, \citenamefont {Harima},\ and\
  \citenamefont {Ōnuki}}]{Shishido05}%
  \BibitemOpen
  \bibfield  {author} {\bibinfo {author} {\bibfnamefont {Hiroaki}\ \bibnamefont
  {Shishido}}, \bibinfo {author} {\bibfnamefont {Rikio}\ \bibnamefont
  {Settai}}, \bibinfo {author} {\bibfnamefont {Hisatomo}\ \bibnamefont
  {Harima}}, \ and\ \bibinfo {author} {\bibfnamefont {Yoshichika}\ \bibnamefont
  {Ōnuki}},\ }\bibfield  {title} {\enquote {\bibinfo {title} {A drastic change
  of the fermi surface at a critical pressure in cerhin5: dhva study under
  pressure},}\ }\href {\doibase 10.1143/JPSJ.74.1103} {\bibfield  {journal}
  {\bibinfo  {journal} {Journal of the Physical Society of Japan}\ }\textbf
  {\bibinfo {volume} {74}},\ \bibinfo {pages} {1103--1106} (\bibinfo {year}
  {2005})}\BibitemShut {NoStop}%
\bibitem [{\citenamefont {Keimer}\ \emph {et~al.}(2015)\citenamefont {Keimer},
  \citenamefont {Kivelson}, \citenamefont {Norman}, \citenamefont {Uchida},\
  and\ \citenamefont {Zaanen}}]{Keimer15}%
  \BibitemOpen
  \bibfield  {author} {\bibinfo {author} {\bibfnamefont {B.}~\bibnamefont
  {Keimer}}, \bibinfo {author} {\bibfnamefont {S.~A.}\ \bibnamefont
  {Kivelson}}, \bibinfo {author} {\bibfnamefont {M.~R.}\ \bibnamefont
  {Norman}}, \bibinfo {author} {\bibfnamefont {S.}~\bibnamefont {Uchida}}, \
  and\ \bibinfo {author} {\bibfnamefont {J.}~\bibnamefont {Zaanen}},\
  }\bibfield  {title} {\enquote {\bibinfo {title} {From quantum matter to
  high-temperature superconductivity in copper oxides},}\ }\href
  {http://dx.doi.org/10.1038/nature14165} {\bibfield  {journal} {\bibinfo
  {journal} {Nature}\ }\textbf {\bibinfo {volume} {518}},\ \bibinfo {pages}
  {179--186} (\bibinfo {year} {2015})}\BibitemShut {NoStop}%
\bibitem [{\citenamefont {Badoux}\ \emph {et~al.}(2016)\citenamefont {Badoux},
  \citenamefont {Tabis}, \citenamefont {Lalibert{\'e}}, \citenamefont
  {Grissonnanche}, \citenamefont {Vignolle}, \citenamefont {Vignolles},
  \citenamefont {B{\'e}ard}, \citenamefont {Bonn}, \citenamefont {Hardy},
  \citenamefont {Liang}, \citenamefont {Doiron-Leyraud}, \citenamefont
  {Taillefer},\ and\ \citenamefont {Proust}}]{Badoux16}%
  \BibitemOpen
  \bibfield  {author} {\bibinfo {author} {\bibfnamefont {S.}~\bibnamefont
  {Badoux}}, \bibinfo {author} {\bibfnamefont {W.}~\bibnamefont {Tabis}},
  \bibinfo {author} {\bibfnamefont {F.}~\bibnamefont {Lalibert{\'e}}}, \bibinfo
  {author} {\bibfnamefont {G.}~\bibnamefont {Grissonnanche}}, \bibinfo {author}
  {\bibfnamefont {B.}~\bibnamefont {Vignolle}}, \bibinfo {author}
  {\bibfnamefont {D.}~\bibnamefont {Vignolles}}, \bibinfo {author}
  {\bibfnamefont {J.}~\bibnamefont {B{\'e}ard}}, \bibinfo {author}
  {\bibfnamefont {D.~A.}\ \bibnamefont {Bonn}}, \bibinfo {author}
  {\bibfnamefont {W.~N.}\ \bibnamefont {Hardy}}, \bibinfo {author}
  {\bibfnamefont {R.}~\bibnamefont {Liang}}, \bibinfo {author} {\bibfnamefont
  {N.}~\bibnamefont {Doiron-Leyraud}}, \bibinfo {author} {\bibfnamefont
  {Louis}\ \bibnamefont {Taillefer}}, \ and\ \bibinfo {author} {\bibfnamefont
  {Cyril}\ \bibnamefont {Proust}},\ }\bibfield  {title} {\enquote {\bibinfo
  {title} {Change of carrier density at the pseudogap critical point of a
  cuprate superconductor},}\ }\href {\doibase 10.1038/nature16983} {\bibfield
  {journal} {\bibinfo  {journal} {Nature}\ }\textbf {\bibinfo {volume} {531}},\
  \bibinfo {pages} {210--214} (\bibinfo {year} {2016})}\BibitemShut {NoStop}%
\bibitem [{\citenamefont {{Alford}}\ \emph {et~al.}(2008)\citenamefont
  {{Alford}}, \citenamefont {{Schmitt}}, \citenamefont {{Rajagopal}},\ and\
  \citenamefont {{Sch{\"a}fer}}}]{Alford2007}%
  \BibitemOpen
  \bibfield  {author} {\bibinfo {author} {\bibfnamefont {Mark~G.}\ \bibnamefont
  {{Alford}}}, \bibinfo {author} {\bibfnamefont {Andreas}\ \bibnamefont
  {{Schmitt}}}, \bibinfo {author} {\bibfnamefont {Krishna}\ \bibnamefont
  {{Rajagopal}}}, \ and\ \bibinfo {author} {\bibfnamefont {Thomas}\
  \bibnamefont {{Sch{\"a}fer}}},\ }\bibfield  {title} {\enquote {\bibinfo
  {title} {{Color superconductivity in dense quark matter}},}\ }\href {\doibase
  10.1103/RevModPhys.80.1455} {\bibfield  {journal} {\bibinfo  {journal}
  {Reviews of Modern Physics}\ }\textbf {\bibinfo {volume} {80}},\ \bibinfo
  {pages} {1455--1515} (\bibinfo {year} {2008})},\ \Eprint
  {http://arxiv.org/abs/0709.4635} {arXiv:0709.4635 [hep-ph]} \BibitemShut
  {NoStop}%
\bibitem [{\citenamefont {Hunt}\ \emph {et~al.}(2013)\citenamefont {Hunt},
  \citenamefont {Sanchez-Yamagishi}, \citenamefont {Young}, \citenamefont
  {Yankowitz}, \citenamefont {LeRoy}, \citenamefont {Watanabe}, \citenamefont
  {Taniguchi}, \citenamefont {Moon}, \citenamefont {Koshino}, \citenamefont
  {Jarillo-Herrero},\ and\ \citenamefont {Ashoori}}]{PJH13}%
  \BibitemOpen
  \bibfield  {author} {\bibinfo {author} {\bibfnamefont {B.}~\bibnamefont
  {Hunt}}, \bibinfo {author} {\bibfnamefont {J.~D.}\ \bibnamefont
  {Sanchez-Yamagishi}}, \bibinfo {author} {\bibfnamefont {A.~F.}\ \bibnamefont
  {Young}}, \bibinfo {author} {\bibfnamefont {M.}~\bibnamefont {Yankowitz}},
  \bibinfo {author} {\bibfnamefont {B.~J.}\ \bibnamefont {LeRoy}}, \bibinfo
  {author} {\bibfnamefont {K.}~\bibnamefont {Watanabe}}, \bibinfo {author}
  {\bibfnamefont {T.}~\bibnamefont {Taniguchi}}, \bibinfo {author}
  {\bibfnamefont {P.}~\bibnamefont {Moon}}, \bibinfo {author} {\bibfnamefont
  {M.}~\bibnamefont {Koshino}}, \bibinfo {author} {\bibfnamefont
  {P.}~\bibnamefont {Jarillo-Herrero}}, \ and\ \bibinfo {author} {\bibfnamefont
  {R.~C.}\ \bibnamefont {Ashoori}},\ }\bibfield  {title} {\enquote {\bibinfo
  {title} {Massive dirac fermions and hofstadter butterfly in a van der waals
  heterostructure},}\ }\href {\doibase 10.1126/science.1237240} {\bibfield
  {journal} {\bibinfo  {journal} {Science}\ }\textbf {\bibinfo {volume}
  {340}},\ \bibinfo {pages} {1427--1430} (\bibinfo {year} {2013})}\BibitemShut
  {NoStop}%
\bibitem [{\citenamefont {Ponomarenko}\ \emph {et~al.}(2013)\citenamefont
  {Ponomarenko}, \citenamefont {Gorbachev}, \citenamefont {Yu}, \citenamefont
  {Elias}, \citenamefont {Jalil}, \citenamefont {Patel}, \citenamefont
  {Mishchenko}, \citenamefont {Mayorov}, \citenamefont {Woods}, \citenamefont
  {Wallbank}, \citenamefont {Mucha-Kruczynski}, \citenamefont {Piot},
  \citenamefont {Potemski}, \citenamefont {Grigorieva}, \citenamefont
  {Novoselov}, \citenamefont {Guinea}, \citenamefont {Fal'ko},\ and\
  \citenamefont {Geim}}]{AKG13}%
  \BibitemOpen
  \bibfield  {author} {\bibinfo {author} {\bibfnamefont {L.~A.}\ \bibnamefont
  {Ponomarenko}}, \bibinfo {author} {\bibfnamefont {R.~V.}\ \bibnamefont
  {Gorbachev}}, \bibinfo {author} {\bibfnamefont {G.~L.}\ \bibnamefont {Yu}},
  \bibinfo {author} {\bibfnamefont {D.~C.}\ \bibnamefont {Elias}}, \bibinfo
  {author} {\bibfnamefont {R.}~\bibnamefont {Jalil}}, \bibinfo {author}
  {\bibfnamefont {A.~A.}\ \bibnamefont {Patel}}, \bibinfo {author}
  {\bibfnamefont {A.}~\bibnamefont {Mishchenko}}, \bibinfo {author}
  {\bibfnamefont {A.~S.}\ \bibnamefont {Mayorov}}, \bibinfo {author}
  {\bibfnamefont {C.~R.}\ \bibnamefont {Woods}}, \bibinfo {author}
  {\bibfnamefont {J.~R.}\ \bibnamefont {Wallbank}}, \bibinfo {author}
  {\bibfnamefont {M.}~\bibnamefont {Mucha-Kruczynski}}, \bibinfo {author}
  {\bibfnamefont {B.~A.}\ \bibnamefont {Piot}}, \bibinfo {author}
  {\bibfnamefont {M.}~\bibnamefont {Potemski}}, \bibinfo {author}
  {\bibfnamefont {I.~V.}\ \bibnamefont {Grigorieva}}, \bibinfo {author}
  {\bibfnamefont {K.~S.}\ \bibnamefont {Novoselov}}, \bibinfo {author}
  {\bibfnamefont {F.}~\bibnamefont {Guinea}}, \bibinfo {author} {\bibfnamefont
  {V.~I.}\ \bibnamefont {Fal'ko}}, \ and\ \bibinfo {author} {\bibfnamefont
  {A.~K.}\ \bibnamefont {Geim}},\ }\bibfield  {title} {\enquote {\bibinfo
  {title} {Cloning of dirac fermions in graphene superlattices},}\ }\href
  {\doibase 10.1038/nature12187} {\bibfield  {journal} {\bibinfo  {journal}
  {Nature}\ }\textbf {\bibinfo {volume} {497}},\ \bibinfo {pages} {594--597}
  (\bibinfo {year} {2013})}\BibitemShut {NoStop}%
\bibitem [{\citenamefont {Dean}\ \emph {et~al.}(2013)\citenamefont {Dean},
  \citenamefont {Wang}, \citenamefont {Maher}, \citenamefont {Forsythe},
  \citenamefont {Ghahari}, \citenamefont {Gao}, \citenamefont {Katoch},
  \citenamefont {Ishigami}, \citenamefont {Moon}, \citenamefont {Koshino},
  \citenamefont {Taniguchi}, \citenamefont {Watanabe}, \citenamefont {Shepard},
  \citenamefont {Hone},\ and\ \citenamefont {Kim}}]{PK13}%
  \BibitemOpen
  \bibfield  {author} {\bibinfo {author} {\bibfnamefont {C.~R.}\ \bibnamefont
  {Dean}}, \bibinfo {author} {\bibfnamefont {L.}~\bibnamefont {Wang}}, \bibinfo
  {author} {\bibfnamefont {P.}~\bibnamefont {Maher}}, \bibinfo {author}
  {\bibfnamefont {C.}~\bibnamefont {Forsythe}}, \bibinfo {author}
  {\bibfnamefont {F.}~\bibnamefont {Ghahari}}, \bibinfo {author} {\bibfnamefont
  {Y.}~\bibnamefont {Gao}}, \bibinfo {author} {\bibfnamefont {J.}~\bibnamefont
  {Katoch}}, \bibinfo {author} {\bibfnamefont {M.}~\bibnamefont {Ishigami}},
  \bibinfo {author} {\bibfnamefont {P.}~\bibnamefont {Moon}}, \bibinfo {author}
  {\bibfnamefont {M.}~\bibnamefont {Koshino}}, \bibinfo {author} {\bibfnamefont
  {T.}~\bibnamefont {Taniguchi}}, \bibinfo {author} {\bibfnamefont
  {K.}~\bibnamefont {Watanabe}}, \bibinfo {author} {\bibfnamefont {K.~L.}\
  \bibnamefont {Shepard}}, \bibinfo {author} {\bibfnamefont {J.}~\bibnamefont
  {Hone}}, \ and\ \bibinfo {author} {\bibfnamefont {P.}~\bibnamefont {Kim}},\
  }\bibfield  {title} {\enquote {\bibinfo {title} {Hofstadter's butterfly and
  the fractal quantum hall effect in moir{\'e} superlattices},}\ }\href
  {\doibase 10.1038/nature12186} {\bibfield  {journal} {\bibinfo  {journal}
  {Nature}\ }\textbf {\bibinfo {volume} {497}},\ \bibinfo {pages} {598--602}
  (\bibinfo {year} {2013})}\BibitemShut {NoStop}%
\bibitem [{\citenamefont {Spanton}\ \emph {et~al.}(2018)\citenamefont
  {Spanton}, \citenamefont {Zibrov}, \citenamefont {Zhou}, \citenamefont
  {Taniguchi}, \citenamefont {Watanabe}, \citenamefont {Zaletel},\ and\
  \citenamefont {Young}}]{AFY18}%
  \BibitemOpen
  \bibfield  {author} {\bibinfo {author} {\bibfnamefont {Eric~M.}\ \bibnamefont
  {Spanton}}, \bibinfo {author} {\bibfnamefont {Alexander~A.}\ \bibnamefont
  {Zibrov}}, \bibinfo {author} {\bibfnamefont {Haoxin}\ \bibnamefont {Zhou}},
  \bibinfo {author} {\bibfnamefont {Takashi}\ \bibnamefont {Taniguchi}},
  \bibinfo {author} {\bibfnamefont {Kenji}\ \bibnamefont {Watanabe}}, \bibinfo
  {author} {\bibfnamefont {Michael~P.}\ \bibnamefont {Zaletel}}, \ and\
  \bibinfo {author} {\bibfnamefont {Andrea~F.}\ \bibnamefont {Young}},\
  }\bibfield  {title} {\enquote {\bibinfo {title} {Observation of fractional
  chern insulators in a van der waals heterostructure},}\ }\href {\doibase
  10.1126/science.aan8458} {\bibfield  {journal} {\bibinfo  {journal}
  {Science}\ }\textbf {\bibinfo {volume} {360}},\ \bibinfo {pages} {62--66}
  (\bibinfo {year} {2018})}\BibitemShut {NoStop}%
\bibitem [{\citenamefont {{Barkeshli}}\ and\ \citenamefont
  {{McGreevy}}(2012)}]{Barkeshli2012}%
  \BibitemOpen
  \bibfield  {author} {\bibinfo {author} {\bibfnamefont {Maissam}\ \bibnamefont
  {{Barkeshli}}}\ and\ \bibinfo {author} {\bibfnamefont {John}\ \bibnamefont
  {{McGreevy}}},\ }\bibfield  {title} {\enquote {\bibinfo {title} {{Continuous
  transitions between composite Fermi liquid and Landau Fermi liquid: A route
  to fractionalized Mott insulators}},}\ }\href {\doibase
  10.1103/PhysRevB.86.075136} {\bibfield  {journal} {\bibinfo  {journal}
  {\prb}\ }\textbf {\bibinfo {volume} {86}},\ \bibinfo {eid} {075136} (\bibinfo
  {year} {2012})},\ \Eprint {http://arxiv.org/abs/1206.6530} {arXiv:1206.6530
  [cond-mat.str-el]} \BibitemShut {NoStop}%
\bibitem [{\citenamefont {Halperin}\ \emph {et~al.}(1993)\citenamefont
  {Halperin}, \citenamefont {Lee},\ and\ \citenamefont {Read}}]{HLR}%
  \BibitemOpen
  \bibfield  {author} {\bibinfo {author} {\bibfnamefont {B.~I.}\ \bibnamefont
  {Halperin}}, \bibinfo {author} {\bibfnamefont {Patrick~A.}\ \bibnamefont
  {Lee}}, \ and\ \bibinfo {author} {\bibfnamefont {Nicholas}\ \bibnamefont
  {Read}},\ }\bibfield  {title} {\enquote {\bibinfo {title} {Theory of the
  half-filled landau level},}\ }\href {\doibase 10.1103/PhysRevB.47.7312}
  {\bibfield  {journal} {\bibinfo  {journal} {Phys. Rev. B}\ }\textbf {\bibinfo
  {volume} {47}},\ \bibinfo {pages} {7312--7343} (\bibinfo {year}
  {1993})}\BibitemShut {NoStop}%
\bibitem [{\citenamefont {{Lee}}\ \emph {et~al.}(2018)\citenamefont {{Lee}},
  \citenamefont {{Wang}}, \citenamefont {{Zaletel}}, \citenamefont
  {{Vishwanath}},\ and\ \citenamefont {{He}}}]{Lee2018}%
  \BibitemOpen
  \bibfield  {author} {\bibinfo {author} {\bibfnamefont {Jong~Yeon}\
  \bibnamefont {{Lee}}}, \bibinfo {author} {\bibfnamefont {Chong}\ \bibnamefont
  {{Wang}}}, \bibinfo {author} {\bibfnamefont {Michael~P.}\ \bibnamefont
  {{Zaletel}}}, \bibinfo {author} {\bibfnamefont {Ashvin}\ \bibnamefont
  {{Vishwanath}}}, \ and\ \bibinfo {author} {\bibfnamefont {Yin-Chen}\
  \bibnamefont {{He}}},\ }\bibfield  {title} {\enquote {\bibinfo {title}
  {{Emergent Multi-Flavor QED$_{3}$ at the Plateau Transition between
  Fractional Chern Insulators: Applications to Graphene Heterostructures}},}\
  }\href {\doibase 10.1103/PhysRevX.8.031015} {\bibfield  {journal} {\bibinfo
  {journal} {Physical Review X}\ }\textbf {\bibinfo {volume} {8}},\ \bibinfo
  {eid} {031015} (\bibinfo {year} {2018})},\ \Eprint
  {http://arxiv.org/abs/1802.09538} {arXiv:1802.09538 [cond-mat.str-el]}
  \BibitemShut {NoStop}%
\bibitem [{\citenamefont {{Bonesteel}}\ \emph {et~al.}(1996)\citenamefont
  {{Bonesteel}}, \citenamefont {{McDonald}},\ and\ \citenamefont
  {{Nayak}}}]{Bonesteel1996}%
  \BibitemOpen
  \bibfield  {author} {\bibinfo {author} {\bibfnamefont {N.~E.}\ \bibnamefont
  {{Bonesteel}}}, \bibinfo {author} {\bibfnamefont {I.~A.}\ \bibnamefont
  {{McDonald}}}, \ and\ \bibinfo {author} {\bibfnamefont {C.}~\bibnamefont
  {{Nayak}}},\ }\bibfield  {title} {\enquote {\bibinfo {title} {{Gauge Fields
  and Pairing in Double-Layer Composite Fermion Metals}},}\ }\href {\doibase
  10.1103/PhysRevLett.77.3009} {\bibfield  {journal} {\bibinfo  {journal}
  {\prl}\ }\textbf {\bibinfo {volume} {77}},\ \bibinfo {pages} {3009--3012}
  (\bibinfo {year} {1996})},\ \Eprint {http://arxiv.org/abs/cond-mat/9601112}
  {arXiv:cond-mat/9601112 [cond-mat]} \BibitemShut {NoStop}%
\bibitem [{\citenamefont {{Sodemann}}\ \emph {et~al.}(2017)\citenamefont
  {{Sodemann}}, \citenamefont {{Kimchi}}, \citenamefont {{Wang}},\ and\
  \citenamefont {{Senthil}}}]{Sodemann2016}%
  \BibitemOpen
  \bibfield  {author} {\bibinfo {author} {\bibfnamefont {Inti}\ \bibnamefont
  {{Sodemann}}}, \bibinfo {author} {\bibfnamefont {Itamar}\ \bibnamefont
  {{Kimchi}}}, \bibinfo {author} {\bibfnamefont {Chong}\ \bibnamefont
  {{Wang}}}, \ and\ \bibinfo {author} {\bibfnamefont {T.}~\bibnamefont
  {{Senthil}}},\ }\bibfield  {title} {\enquote {\bibinfo {title} {{Composite
  fermion duality for half-filled multicomponent Landau levels}},}\ }\href
  {\doibase 10.1103/PhysRevB.95.085135} {\bibfield  {journal} {\bibinfo
  {journal} {\prb}\ }\textbf {\bibinfo {volume} {95}},\ \bibinfo {eid} {085135}
  (\bibinfo {year} {2017})},\ \Eprint {http://arxiv.org/abs/1609.08616}
  {arXiv:1609.08616 [cond-mat.str-el]} \BibitemShut {NoStop}%
\bibitem [{\citenamefont {{Isobe}}\ and\ \citenamefont
  {{Fu}}(2017)}]{Isobe2016}%
  \BibitemOpen
  \bibfield  {author} {\bibinfo {author} {\bibfnamefont {Hiroki}\ \bibnamefont
  {{Isobe}}}\ and\ \bibinfo {author} {\bibfnamefont {Liang}\ \bibnamefont
  {{Fu}}},\ }\bibfield  {title} {\enquote {\bibinfo {title} {{Interlayer
  Pairing Symmetry of Composite Fermions in Quantum Hall Bilayers}},}\ }\href
  {\doibase 10.1103/PhysRevLett.118.166401} {\bibfield  {journal} {\bibinfo
  {journal} {\prl}\ }\textbf {\bibinfo {volume} {118}},\ \bibinfo {eid}
  {166401} (\bibinfo {year} {2017})},\ \Eprint
  {http://arxiv.org/abs/1609.09063} {arXiv:1609.09063 [cond-mat.str-el]}
  \BibitemShut {NoStop}%
\bibitem [{sup(2020)}]{supp}%
  \BibitemOpen
  \href@noop {} {} (\bibinfo {year} {2020}),\ \bibinfo {note} {see
  supplementary material for additional details.}\BibitemShut {Stop}%
\bibitem [{\citenamefont {{Senthil}}(2008{\natexlab{b}})}]{Senthil2008a}%
  \BibitemOpen
  \bibfield  {author} {\bibinfo {author} {\bibfnamefont {T.}~\bibnamefont
  {{Senthil}}},\ }\bibfield  {title} {\enquote {\bibinfo {title} {{Critical
  Fermi surfaces and non-Fermi liquid metals}},}\ }\href {\doibase
  10.1103/PhysRevB.78.035103} {\bibfield  {journal} {\bibinfo  {journal}
  {\prb}\ }\textbf {\bibinfo {volume} {78}},\ \bibinfo {eid} {035103} (\bibinfo
  {year} {2008}{\natexlab{b}})},\ \Eprint {http://arxiv.org/abs/0803.4009}
  {arXiv:0803.4009 [cond-mat.str-el]} \BibitemShut {NoStop}%
\bibitem [{\citenamefont {Wen}(2004)}]{Wen2004Book}%
  \BibitemOpen
  \bibfield  {author} {\bibinfo {author} {\bibfnamefont {Xiao-Gang}\
  \bibnamefont {Wen}},\ }\href@noop {} {\emph {\bibinfo {title} {Quantum field
  theory of many-body systems}}}\ (\bibinfo  {publisher} {Oxford University
  Press, Oxford},\ \bibinfo {year} {2004})\BibitemShut {NoStop}%
\bibitem [{\citenamefont {Barkeshli}\ and\ \citenamefont
  {McGreevy}(2014)}]{Barkeshli2012a}%
  \BibitemOpen
  \bibfield  {author} {\bibinfo {author} {\bibfnamefont {Maissam}\ \bibnamefont
  {Barkeshli}}\ and\ \bibinfo {author} {\bibfnamefont {John}\ \bibnamefont
  {McGreevy}},\ }\bibfield  {title} {\enquote {\bibinfo {title} {Continuous
  transition between fractional quantum hall and superfluid states},}\ }\href
  {\doibase 10.1103/PhysRevB.89.235116} {\bibfield  {journal} {\bibinfo
  {journal} {Phys. Rev. B}\ }\textbf {\bibinfo {volume} {89}},\ \bibinfo
  {pages} {235116} (\bibinfo {year} {2014})}\BibitemShut {NoStop}%
\bibitem [{\citenamefont {Zou}\ and\ \citenamefont {He}(2020)}]{Zou2018}%
  \BibitemOpen
  \bibfield  {author} {\bibinfo {author} {\bibfnamefont {Liujun}\ \bibnamefont
  {Zou}}\ and\ \bibinfo {author} {\bibfnamefont {Yin-Chen}\ \bibnamefont
  {He}},\ }\bibfield  {title} {\enquote {\bibinfo {title} {Field-induced
  ${\mathrm{qcd}}_{3}$-chern-simons quantum criticalities in kitaev
  materials},}\ }\href {\doibase 10.1103/PhysRevResearch.2.013072} {\bibfield
  {journal} {\bibinfo  {journal} {Phys. Rev. Research}\ }\textbf {\bibinfo
  {volume} {2}},\ \bibinfo {pages} {013072} (\bibinfo {year}
  {2020})}\BibitemShut {NoStop}%
\bibitem [{\citenamefont {Dasgupta}\ and\ \citenamefont
  {Halperin}(1981)}]{Dasgupta1981}%
  \BibitemOpen
  \bibfield  {author} {\bibinfo {author} {\bibfnamefont {C.}~\bibnamefont
  {Dasgupta}}\ and\ \bibinfo {author} {\bibfnamefont {B.~I.}\ \bibnamefont
  {Halperin}},\ }\bibfield  {title} {\enquote {\bibinfo {title} {Phase
  transition in a lattice model of superconductivity},}\ }\href {\doibase
  10.1103/PhysRevLett.47.1556} {\bibfield  {journal} {\bibinfo  {journal}
  {Phys. Rev. Lett.}\ }\textbf {\bibinfo {volume} {47}},\ \bibinfo {pages}
  {1556--1560} (\bibinfo {year} {1981})}\BibitemShut {NoStop}%
\bibitem [{\citenamefont {{Chen}}\ \emph {et~al.}(1993)\citenamefont {{Chen}},
  \citenamefont {{Fisher}},\ and\ \citenamefont {{Wu}}}]{Chen1993}%
  \BibitemOpen
  \bibfield  {author} {\bibinfo {author} {\bibfnamefont {Wei}\ \bibnamefont
  {{Chen}}}, \bibinfo {author} {\bibfnamefont {Matthew P.~A.}\ \bibnamefont
  {{Fisher}}}, \ and\ \bibinfo {author} {\bibfnamefont {Yong-Shi}\ \bibnamefont
  {{Wu}}},\ }\bibfield  {title} {\enquote {\bibinfo {title} {{Mott transition
  in an anyon gas}},}\ }\href {\doibase 10.1103/PhysRevB.48.13749} {\bibfield
  {journal} {\bibinfo  {journal} {\prb}\ }\textbf {\bibinfo {volume} {48}},\
  \bibinfo {pages} {13749--13761} (\bibinfo {year} {1993})},\ \Eprint
  {http://arxiv.org/abs/cond-mat/9301037} {arXiv:cond-mat/9301037 [cond-mat]}
  \BibitemShut {NoStop}%
\bibitem [{\citenamefont {{Senthil}}\ and\ \citenamefont
  {{Levin}}(2013)}]{Senthil2012}%
  \BibitemOpen
  \bibfield  {author} {\bibinfo {author} {\bibfnamefont {T.}~\bibnamefont
  {{Senthil}}}\ and\ \bibinfo {author} {\bibfnamefont {Michael}\ \bibnamefont
  {{Levin}}},\ }\bibfield  {title} {\enquote {\bibinfo {title} {{Integer
  Quantum Hall Effect for Bosons}},}\ }\href {\doibase
  10.1103/PhysRevLett.110.046801} {\bibfield  {journal} {\bibinfo  {journal}
  {\prl}\ }\textbf {\bibinfo {volume} {110}},\ \bibinfo {eid} {046801}
  (\bibinfo {year} {2013})},\ \Eprint {http://arxiv.org/abs/1206.1604}
  {arXiv:1206.1604 [cond-mat.str-el]} \BibitemShut {NoStop}%
\bibitem [{\citenamefont {{Zou}}\ and\ \citenamefont
  {{Senthil}}(2016)}]{Zou2016}%
  \BibitemOpen
  \bibfield  {author} {\bibinfo {author} {\bibfnamefont {Liujun}\ \bibnamefont
  {{Zou}}}\ and\ \bibinfo {author} {\bibfnamefont {T.}~\bibnamefont
  {{Senthil}}},\ }\bibfield  {title} {\enquote {\bibinfo {title} {{Dimensional
  decoupling at continuous quantum critical Mott transitions}},}\ }\href
  {\doibase 10.1103/PhysRevB.94.115113} {\bibfield  {journal} {\bibinfo
  {journal} {\prb}\ }\textbf {\bibinfo {volume} {94}},\ \bibinfo {eid} {115113}
  (\bibinfo {year} {2016})},\ \Eprint {http://arxiv.org/abs/1603.09359}
  {arXiv:1603.09359 [cond-mat.str-el]} \BibitemShut {NoStop}%
\bibitem [{\citenamefont {Kohn}\ and\ \citenamefont
  {Luttinger}(1965)}]{Kohn1965}%
  \BibitemOpen
  \bibfield  {author} {\bibinfo {author} {\bibfnamefont {W.}~\bibnamefont
  {Kohn}}\ and\ \bibinfo {author} {\bibfnamefont {J.~M.}\ \bibnamefont
  {Luttinger}},\ }\bibfield  {title} {\enquote {\bibinfo {title} {New mechanism
  for superconductivity},}\ }\href {\doibase 10.1103/PhysRevLett.15.524}
  {\bibfield  {journal} {\bibinfo  {journal} {Phys. Rev. Lett.}\ }\textbf
  {\bibinfo {volume} {15}},\ \bibinfo {pages} {524--526} (\bibinfo {year}
  {1965})}\BibitemShut {NoStop}%
\bibitem [{\citenamefont {{Shankar}}(1994)}]{Shankar1994}%
  \BibitemOpen
  \bibfield  {author} {\bibinfo {author} {\bibfnamefont {R.}~\bibnamefont
  {{Shankar}}},\ }\bibfield  {title} {\enquote {\bibinfo {title}
  {{Renormalization-group approach to interacting fermions}},}\ }\href
  {\doibase 10.1103/RevModPhys.66.129} {\bibfield  {journal} {\bibinfo
  {journal} {Reviews of Modern Physics}\ }\textbf {\bibinfo {volume} {66}},\
  \bibinfo {pages} {129--192} (\bibinfo {year} {1994})},\ \Eprint
  {http://arxiv.org/abs/cond-mat/9307009} {arXiv:cond-mat/9307009 [cond-mat]}
  \BibitemShut {NoStop}%
\bibitem [{\citenamefont {{Kitaev}}(2006)}]{Kitaev2006}%
  \BibitemOpen
  \bibfield  {author} {\bibinfo {author} {\bibfnamefont {Alexei}\ \bibnamefont
  {{Kitaev}}},\ }\bibfield  {title} {\enquote {\bibinfo {title} {{Anyons in an
  exactly solved model and beyond}},}\ }\href {\doibase
  10.1016/j.aop.2005.10.005} {\bibfield  {journal} {\bibinfo  {journal} {Annals
  of Physics}\ }\textbf {\bibinfo {volume} {321}},\ \bibinfo {pages} {2--111}
  (\bibinfo {year} {2006})},\ \Eprint {http://arxiv.org/abs/cond-mat/0506438}
  {arXiv:cond-mat/0506438 [cond-mat.mes-hall]} \BibitemShut {NoStop}%
\bibitem [{\citenamefont {{Senthil}}\ and\ \citenamefont
  {{Fisher}}(2000)}]{Senthil2000}%
  \BibitemOpen
  \bibfield  {author} {\bibinfo {author} {\bibfnamefont {T.}~\bibnamefont
  {{Senthil}}}\ and\ \bibinfo {author} {\bibfnamefont {Matthew P.~A.}\
  \bibnamefont {{Fisher}}},\ }\bibfield  {title} {\enquote {\bibinfo {title}
  {{Z$_{2}$ gauge theory of electron fractionalization in strongly correlated
  systems}},}\ }\href {\doibase 10.1103/PhysRevB.62.7850} {\bibfield  {journal}
  {\bibinfo  {journal} {\prb}\ }\textbf {\bibinfo {volume} {62}},\ \bibinfo
  {pages} {7850--7881} (\bibinfo {year} {2000})},\ \Eprint
  {http://arxiv.org/abs/cond-mat/9910224} {arXiv:cond-mat/9910224
  [cond-mat.str-el]} \BibitemShut {NoStop}%
\bibitem [{\citenamefont {{Ye}}\ and\ \citenamefont
  {{Sachdev}}(1998)}]{Ye1997}%
  \BibitemOpen
  \bibfield  {author} {\bibinfo {author} {\bibfnamefont {Jinwu}\ \bibnamefont
  {{Ye}}}\ and\ \bibinfo {author} {\bibfnamefont {Subir}\ \bibnamefont
  {{Sachdev}}},\ }\bibfield  {title} {\enquote {\bibinfo {title} {{Coulomb
  Interactions at Quantum Hall Critical Points of Systems in a Periodic
  Potential}},}\ }\href {\doibase 10.1103/PhysRevLett.80.5409} {\bibfield
  {journal} {\bibinfo  {journal} {\prl}\ }\textbf {\bibinfo {volume} {80}},\
  \bibinfo {pages} {5409--5412} (\bibinfo {year} {1998})},\ \Eprint
  {http://arxiv.org/abs/cond-mat/9712161} {arXiv:cond-mat/9712161
  [cond-mat.mes-hall]} \BibitemShut {NoStop}%
\bibitem [{\citenamefont {{Cooper}}\ and\ \citenamefont
  {{Dalibard}}(2013)}]{Cooper2013}%
  \BibitemOpen
  \bibfield  {author} {\bibinfo {author} {\bibfnamefont {Nigel~R.}\
  \bibnamefont {{Cooper}}}\ and\ \bibinfo {author} {\bibfnamefont {Jean}\
  \bibnamefont {{Dalibard}}},\ }\bibfield  {title} {\enquote {\bibinfo {title}
  {{Reaching Fractional Quantum Hall States with Optical Flux Lattices}},}\
  }\href {\doibase 10.1103/PhysRevLett.110.185301} {\bibfield  {journal}
  {\bibinfo  {journal} {\prl}\ }\textbf {\bibinfo {volume} {110}},\ \bibinfo
  {eid} {185301} (\bibinfo {year} {2013})},\ \Eprint
  {http://arxiv.org/abs/1212.3552} {arXiv:1212.3552 [cond-mat.quant-gas]}
  \BibitemShut {NoStop}%
\bibitem [{\citenamefont {{Yao}}\ \emph {et~al.}(2013)\citenamefont {{Yao}},
  \citenamefont {{Gorshkov}}, \citenamefont {{Laumann}}, \citenamefont
  {{L{\"a}uchli}}, \citenamefont {{Ye}},\ and\ \citenamefont
  {{Lukin}}}]{Yao2013}%
  \BibitemOpen
  \bibfield  {author} {\bibinfo {author} {\bibfnamefont {N.~Y.}\ \bibnamefont
  {{Yao}}}, \bibinfo {author} {\bibfnamefont {A.~V.}\ \bibnamefont
  {{Gorshkov}}}, \bibinfo {author} {\bibfnamefont {C.~R.}\ \bibnamefont
  {{Laumann}}}, \bibinfo {author} {\bibfnamefont {A.~M.}\ \bibnamefont
  {{L{\"a}uchli}}}, \bibinfo {author} {\bibfnamefont {J.}~\bibnamefont {{Ye}}},
  \ and\ \bibinfo {author} {\bibfnamefont {M.~D.}\ \bibnamefont {{Lukin}}},\
  }\bibfield  {title} {\enquote {\bibinfo {title} {{Realizing Fractional Chern
  Insulators in Dipolar Spin Systems}},}\ }\href {\doibase
  10.1103/PhysRevLett.110.185302} {\bibfield  {journal} {\bibinfo  {journal}
  {\prl}\ }\textbf {\bibinfo {volume} {110}},\ \bibinfo {eid} {185302}
  (\bibinfo {year} {2013})},\ \Eprint {http://arxiv.org/abs/1212.4839}
  {arXiv:1212.4839 [cond-mat.str-el]} \BibitemShut {NoStop}%
\bibitem [{\citenamefont {{Borokhov}}\ \emph {et~al.}(2002)\citenamefont
  {{Borokhov}}, \citenamefont {{Kapustin}},\ and\ \citenamefont
  {{Wu}}}]{Borokhov2002}%
  \BibitemOpen
  \bibfield  {author} {\bibinfo {author} {\bibfnamefont {Vadim}\ \bibnamefont
  {{Borokhov}}}, \bibinfo {author} {\bibfnamefont {Anton}\ \bibnamefont
  {{Kapustin}}}, \ and\ \bibinfo {author} {\bibfnamefont {Xinkai}\ \bibnamefont
  {{Wu}}},\ }\bibfield  {title} {\enquote {\bibinfo {title} {{Topological
  Disorder Operators in Three-Dimensional Conformal Field Theory}},}\ }\href
  {\doibase 10.1088/1126-6708/2002/11/049} {\bibfield  {journal} {\bibinfo
  {journal} {Journal of High Energy Physics}\ }\textbf {\bibinfo {volume}
  {2002}},\ \bibinfo {eid} {049} (\bibinfo {year} {2002})},\ \Eprint
  {http://arxiv.org/abs/hep-th/0206054} {arXiv:hep-th/0206054 [hep-th]}
  \BibitemShut {NoStop}%
\bibitem [{\citenamefont {{Nandkishore}}\ \emph {et~al.}(2012)\citenamefont
  {{Nandkishore}}, \citenamefont {{Metlitski}},\ and\ \citenamefont
  {{Senthil}}}]{Nandkishore2012}%
  \BibitemOpen
  \bibfield  {author} {\bibinfo {author} {\bibfnamefont {Rahul}\ \bibnamefont
  {{Nandkishore}}}, \bibinfo {author} {\bibfnamefont {Max~A.}\ \bibnamefont
  {{Metlitski}}}, \ and\ \bibinfo {author} {\bibfnamefont {T.}~\bibnamefont
  {{Senthil}}},\ }\bibfield  {title} {\enquote {\bibinfo {title} {{Orthogonal
  metals: The simplest non-Fermi liquids}},}\ }\href {\doibase
  10.1103/PhysRevB.86.045128} {\bibfield  {journal} {\bibinfo  {journal}
  {\prb}\ }\textbf {\bibinfo {volume} {86}},\ \bibinfo {eid} {045128} (\bibinfo
  {year} {2012})},\ \Eprint {http://arxiv.org/abs/1201.5998} {arXiv:1201.5998
  [cond-mat.str-el]} \BibitemShut {NoStop}%
\bibitem [{\citenamefont {He}\ \emph {et~al.}(1993)\citenamefont {He},
  \citenamefont {Platzman},\ and\ \citenamefont {Halperin}}]{BH93}%
  \BibitemOpen
  \bibfield  {author} {\bibinfo {author} {\bibfnamefont {Song}\ \bibnamefont
  {He}}, \bibinfo {author} {\bibfnamefont {P.~M.}\ \bibnamefont {Platzman}}, \
  and\ \bibinfo {author} {\bibfnamefont {B.~I.}\ \bibnamefont {Halperin}},\
  }\bibfield  {title} {\enquote {\bibinfo {title} {Tunneling into a
  two-dimensional electron system in a strong magnetic field},}\ }\href
  {\doibase 10.1103/PhysRevLett.71.777} {\bibfield  {journal} {\bibinfo
  {journal} {Phys. Rev. Lett.}\ }\textbf {\bibinfo {volume} {71}},\ \bibinfo
  {pages} {777--780} (\bibinfo {year} {1993})}\BibitemShut {NoStop}%
\bibitem [{\citenamefont {{Kim}}\ and\ \citenamefont {{Wen}}(1994)}]{Kim1994}%
  \BibitemOpen
  \bibfield  {author} {\bibinfo {author} {\bibfnamefont {Yong~Baek}\
  \bibnamefont {{Kim}}}\ and\ \bibinfo {author} {\bibfnamefont {Xiao-Gang}\
  \bibnamefont {{Wen}}},\ }\bibfield  {title} {\enquote {\bibinfo {title}
  {{Instantons and the spectral function of electrons in the half-filled Landau
  level}},}\ }\href {\doibase 10.1103/PhysRevB.50.8078} {\bibfield  {journal}
  {\bibinfo  {journal} {\prb}\ }\textbf {\bibinfo {volume} {50}},\ \bibinfo
  {pages} {8078--8081} (\bibinfo {year} {1994})},\ \Eprint
  {http://arxiv.org/abs/cond-mat/9401032} {arXiv:cond-mat/9401032 [cond-mat]}
  \BibitemShut {NoStop}%
\bibitem [{\citenamefont {Ioffe}\ and\ \citenamefont
  {Larkin}(1989)}]{Ioffe1989}%
  \BibitemOpen
  \bibfield  {author} {\bibinfo {author} {\bibfnamefont {L.~B.}\ \bibnamefont
  {Ioffe}}\ and\ \bibinfo {author} {\bibfnamefont {A.~I.}\ \bibnamefont
  {Larkin}},\ }\bibfield  {title} {\enquote {\bibinfo {title} {Gapless fermions
  and gauge fields in dielectrics},}\ }\href {\doibase
  10.1103/PhysRevB.39.8988} {\bibfield  {journal} {\bibinfo  {journal} {Phys.
  Rev. B}\ }\textbf {\bibinfo {volume} {39}},\ \bibinfo {pages} {8988--8999}
  (\bibinfo {year} {1989})}\BibitemShut {NoStop}%
\bibitem [{\citenamefont {{Zhang}}\ and\ \citenamefont
  {{Senthil}}(2020)}]{Zhang2020}%
  \BibitemOpen
  \bibfield  {author} {\bibinfo {author} {\bibfnamefont {Ya-Hui}\ \bibnamefont
  {{Zhang}}}\ and\ \bibinfo {author} {\bibfnamefont {T.}~\bibnamefont
  {{Senthil}}},\ }\bibfield  {title} {\enquote {\bibinfo {title} {{Quantum Hall
  spin liquids and their possible realization in moir{\'e} systems}},}\
  }\href@noop {} {\bibfield  {journal} {\bibinfo  {journal} {arXiv e-prints}\
  ,\ \bibinfo {eid} {arXiv:2003.13702}} (\bibinfo {year} {2020})},\ \Eprint
  {http://arxiv.org/abs/2003.13702} {arXiv:2003.13702 [cond-mat.str-el]}
  \BibitemShut {NoStop}%
\bibitem [{\citenamefont {{Nayak}}\ and\ \citenamefont
  {{Wilczek}}(1994)}]{Nayak1994}%
  \BibitemOpen
  \bibfield  {author} {\bibinfo {author} {\bibfnamefont {Chetan}\ \bibnamefont
  {{Nayak}}}\ and\ \bibinfo {author} {\bibfnamefont {Frank}\ \bibnamefont
  {{Wilczek}}},\ }\bibfield  {title} {\enquote {\bibinfo {title} {{Non-Fermi
  liquid fixed point in 2 + 1 dimensions}},}\ }\href {\doibase
  10.1016/0550-3213(94)90477-4} {\bibfield  {journal} {\bibinfo  {journal}
  {Nuclear Physics B}\ }\textbf {\bibinfo {volume} {417}},\ \bibinfo {pages}
  {359--373} (\bibinfo {year} {1994})},\ \Eprint
  {http://arxiv.org/abs/cond-mat/9312086} {arXiv:cond-mat/9312086 [cond-mat]}
  \BibitemShut {NoStop}%
\bibitem [{\citenamefont {{Lee}}(2008)}]{Lee2008}%
  \BibitemOpen
  \bibfield  {author} {\bibinfo {author} {\bibfnamefont {Sung-Sik}\
  \bibnamefont {{Lee}}},\ }\bibfield  {title} {\enquote {\bibinfo {title}
  {{Stability of the U(1) spin liquid with a spinon Fermi surface in 2+1
  dimensions}},}\ }\href {\doibase 10.1103/PhysRevB.78.085129} {\bibfield
  {journal} {\bibinfo  {journal} {\prb}\ }\textbf {\bibinfo {volume} {78}},\
  \bibinfo {eid} {085129} (\bibinfo {year} {2008})},\ \Eprint
  {http://arxiv.org/abs/0804.3800} {arXiv:0804.3800 [cond-mat.str-el]}
  \BibitemShut {NoStop}%
\bibitem [{\citenamefont {{Lee}}(2009)}]{Lee2009}%
  \BibitemOpen
  \bibfield  {author} {\bibinfo {author} {\bibfnamefont {Sung-Sik}\
  \bibnamefont {{Lee}}},\ }\bibfield  {title} {\enquote {\bibinfo {title}
  {{Low-energy effective theory of Fermi surface coupled with U(1) gauge field
  in 2+1 dimensions}},}\ }\href {\doibase 10.1103/PhysRevB.80.165102}
  {\bibfield  {journal} {\bibinfo  {journal} {\prb}\ }\textbf {\bibinfo
  {volume} {80}},\ \bibinfo {eid} {165102} (\bibinfo {year} {2009})},\ \Eprint
  {http://arxiv.org/abs/0905.4532} {arXiv:0905.4532 [cond-mat.str-el]}
  \BibitemShut {NoStop}%
\bibitem [{\citenamefont {{Metlitski}}\ and\ \citenamefont
  {{Sachdev}}(2010)}]{Metlitski2010}%
  \BibitemOpen
  \bibfield  {author} {\bibinfo {author} {\bibfnamefont {Max~A.}\ \bibnamefont
  {{Metlitski}}}\ and\ \bibinfo {author} {\bibfnamefont {Subir}\ \bibnamefont
  {{Sachdev}}},\ }\bibfield  {title} {\enquote {\bibinfo {title} {{Quantum
  phase transitions of metals in two spatial dimensions. I. Ising-nematic
  order}},}\ }\href {\doibase 10.1103/PhysRevB.82.075127} {\bibfield  {journal}
  {\bibinfo  {journal} {\prb}\ }\textbf {\bibinfo {volume} {82}},\ \bibinfo
  {eid} {075127} (\bibinfo {year} {2010})},\ \Eprint
  {http://arxiv.org/abs/1001.1153} {arXiv:1001.1153 [cond-mat.str-el]}
  \BibitemShut {NoStop}%
\bibitem [{\citenamefont {{Mross}}\ \emph {et~al.}(2010)\citenamefont
  {{Mross}}, \citenamefont {{McGreevy}}, \citenamefont {{Liu}},\ and\
  \citenamefont {{Senthil}}}]{Mross2010}%
  \BibitemOpen
  \bibfield  {author} {\bibinfo {author} {\bibfnamefont {David~F.}\
  \bibnamefont {{Mross}}}, \bibinfo {author} {\bibfnamefont {John}\
  \bibnamefont {{McGreevy}}}, \bibinfo {author} {\bibfnamefont {Hong}\
  \bibnamefont {{Liu}}}, \ and\ \bibinfo {author} {\bibfnamefont
  {T.}~\bibnamefont {{Senthil}}},\ }\bibfield  {title} {\enquote {\bibinfo
  {title} {{Controlled expansion for certain non-Fermi-liquid metals}},}\
  }\href {\doibase 10.1103/PhysRevB.82.045121} {\bibfield  {journal} {\bibinfo
  {journal} {\prb}\ }\textbf {\bibinfo {volume} {82}},\ \bibinfo {eid} {045121}
  (\bibinfo {year} {2010})},\ \Eprint {http://arxiv.org/abs/1003.0894}
  {arXiv:1003.0894 [cond-mat.str-el]} \BibitemShut {NoStop}%
\bibitem [{\citenamefont {Metlitski}\ \emph {et~al.}(2015)\citenamefont
  {Metlitski}, \citenamefont {Mross}, \citenamefont {Sachdev},\ and\
  \citenamefont {Senthil}}]{Metlitski2014}%
  \BibitemOpen
  \bibfield  {author} {\bibinfo {author} {\bibfnamefont {Max~A.}\ \bibnamefont
  {Metlitski}}, \bibinfo {author} {\bibfnamefont {David~F.}\ \bibnamefont
  {Mross}}, \bibinfo {author} {\bibfnamefont {Subir}\ \bibnamefont {Sachdev}},
  \ and\ \bibinfo {author} {\bibfnamefont {T.}~\bibnamefont {Senthil}},\
  }\bibfield  {title} {\enquote {\bibinfo {title} {Cooper pairing in non-fermi
  liquids},}\ }\href {\doibase 10.1103/PhysRevB.91.115111} {\bibfield
  {journal} {\bibinfo  {journal} {Phys. Rev. B}\ }\textbf {\bibinfo {volume}
  {91}},\ \bibinfo {pages} {115111} (\bibinfo {year} {2015})}\BibitemShut
  {NoStop}%
\end{thebibliography}%

\end{document}